\documentclass[]{aa}  

\usepackage{graphicx}
\usepackage{txfonts}
\usepackage[svgnames]{xcolor}
\begin{document} 

\title{Blue supergiants as tests for stellar physics}

   \author{Cyril Georgy\inst{1}
          \and
          Hideyuki Saio\inst{2}
          \and
          Georges Meynet\inst{1}
          }

   \institute{Department of Astronomy, University of Geneva, Chemin Pegasi 51, 1290 Versoix, Switzerland\\
              \email{cyril.georgy@unige.ch}
          \and
    Astronomical Institute, Graduate school of Science, Tohoku  University, 980-8578, Japan
             }
    \authorrunning{Georgy et al.}

   \date{}

\abstract{Massive star evolution is still poorly understood, and observational tests are required to discriminate between different implementations of physical phenomenon in stellar evolution codes.}
{By confronting stellar evolution models with observed properties of blue supergiants, such as pulsations, chemical composition and position in the Hertzsprung-Russell diagram, we aim at determining which of the criterion used for convection (Schwarzschild or Ledoux) is best able to explain the observations.}
{We compute state-of-the-art stellar evolution models with either the Schwarzschild or the Ledoux criterion for convection. Models are for $14$ to $35\,M_\sun$ at solar or Large Magellanic Cloud metallicity. For each model, we compute the pulsation properties to know when radial modes are excited. We then compare our results with the position of blue supergiants in the Hertzsprung-Russell diagram, with their surface chemical composition, and with their variability.}
{Our results at Large Magellanic Cloud metallicity shows only a slight preference for the Ledoux criterion over the Schwarzschild one in reproducing at the same time the observed properties of blue supergiants, even if the Schwarzschild criterion cannot be excluded at this metallicity. We check that changing the overshoot parameter at solar metallicity does not improve the situation. We also check that our models are able to reproduce the position of Galactic blue supergiant in the flux-weighted-gravity -- luminosity relation.}
{We confirm that overall, models computed with the Ledoux criterion are slightly better in matching observations. Our results also support the idea that most Galactic $\alpha$ Cyg variables are blue supergiants of the group 2, i.e. stars that have been through a previous red supergiant phase where they have lost large amount of mass.} 


   \maketitle
%

\section{Introduction}

Although massive stars, progenitors of core-collapse supernovae, play  important roles in the chemical evolution of galaxies, their evolution is still poorly understood. Main uncertainties come from our insufficient knowledge of accurate mass-loss rates at various evolutionary stages, the efficiencies of  rotational mixing and angular momentum transport, and the range of convective/semi-convective mixing including the extent of overshooting. Constraints to theories for these phenomena should be obtained from comparison of theoretical  models with observation. Surface CNO abundances  are very useful for constraining theoretical models of rotational  mixing in massive stars. Comparison of the surface CNO elements of B stars with theoretical models has been discussed actively \citep[e.g.,][]{Hunter2009,Przy2010,Brott2011,Maeder2014,Martins2015a}, but there remain  some difficulties reconciling theoretical and numerical results with observations, even for main-sequence (MS) stars.

Convection is one of the main transport process to be included in stellar evolution calculations. Despite its importance, it is still poorly understood and its treatment in stellar evolution codes is yet based on very simple models, such as the well known ``mixing-length theory'' \citep{Boehm-Vitense1958a}. Moreover, the position of the convective boundaries relies on one of the stability criterion \citep[either the Schwarzschild criterion or the Ledoux one, see e.g.][]{Maeder2009}. It is known since a long time \citep[e.g.][]{Maeder1981c} that using either of these criterion produces too small cores, making mandatory to arbitrary extend the size of the convective cores (this is called the ``overshooting''). How this extension has to be done, and what is the efficiency of the mixing of chemical elements throughout the convective boundary (so called ``convective boundary mixing'') is so far unknown and largely contributes to our lack of understanding of massive star evolution.

During the last decade, intensive efforts have been started to improve the situation by using three-dimension (3d) hydrodynamics simulation of convection in stellar interior conditions \citep[][among others]{Meakin2007a,Freytag2008a,Augustson2012a,viallet2013a,muller2016a,Cristini2017a,Mocak2018a}. Despite the increasing sophistication and numerical resolution of these simulations, and the powerful insight they allow on stellar convection, we are still far from having new and more precise algorithms to deal with this phenomenon. Some attempts are currently made to include in classical stellar evolution codes lessons learned from 3d hydrodynamics simulations \citep{Scott2021a}. To validate these approaches, new observational tests are needed \citep[for example see][]{Tkachenko2020a}.

We have been pursuing to obtain constraints to theories by using pulsationnal properties of blue supergiant (BSG) stars \citep{Saio2013,Georgy2014}.  Blue supergiants are expected to consist of two groups; one group of stars are evolving toward the red supergiant (RSG) stage (group 1 hereafter), the others are returned back from the RSG stage (group 2 hereafter). We distinguish the two groups using the pulsation properties; BSG stars after RSG show evidence of radial pulsation ($\alpha$ Cyg variables), when BSG evolving from the MS to the RSG branch (first crossing) do not \citep{Saio2013}. We interpreted Deneb ($\alpha$ Cyg) and Rigel ($\beta$ Ori) as post RSG according to their pulsations which can be reproduced by models which returned to BSG region after losing significant mass during the RSG phase. One puzzle emerged, however: post-RSG models predict surface N/C and N/O ratios much larger than those of Deneb and Rigel, of which N/C and N/O ratios are rather consistent with those of pre-RSG models. The discrepancy indicates that internal mixing was somewhat too strong in our models. \citet{Georgy2014} proposed a possible solution for the problem; i.e., the problem of the surface CNO abundance ratios might be remedied by using the Ledoux criterion in determining the convective/radiative boundaries. However, the latter work focused on only one mass, namely $25\,M_\sun$, at solar metallicity. In this paper, we intent to extend our previous researches to a broader mass domain and also to the metallicity of the Large Magellanic Cloud (LMC), in order to check if our finding remains valid in other regimes.

Most of the stellar evolution calculations discussed in the literature have been done using Schwarzschild criterion so that discussions of the stellar structure and evolution under the Ledoux criterion are rare in spite that it is not yet understood which criterion should be used. In this paper we first discuss the structure and evolution of massive stars calculated with the Ledoux criterion. Then, we compare observed surface compositions of BSGs with those of models obtained by using the Ledoux or the Schwarzschild criterion. Finally, we revisit the $M_\text{bol}-\log g_F$ (luminosity -- flux-weighted surface gravity) relation of the BSGs.

\section{Massive star models with the Ledoux criterion}
As is well known there are two ways to determine the boundary between convection and radiative regions. If the Schwarzschild criterion is employed, convection is assumed to occur if
\begin{equation}
\left(\frac{\text{d}\ln T}{\text{d}\ln P}\right)_\text{rad} > \left(\frac{\text{d}\ln T}{\text{d}\ln P}\right)_\text{ad},
\label{eq:sch}
\end{equation}
where the temperature gradients with subscripts $_\text{rad}$ and $_\text{ad}$ indicate the temperature gradient realized if all energy is carried by radiative diffusion \citep[see e.g.][for the detailed definitions]{Maeder2009}, and adiabatic temperature gradient, respectively. On the other hand, if we adopt the Ledoux criterion, we assume that the convection occurs if
\begin{equation}
\left(\frac{\text{d}\ln T}{\text{d}\ln P}\right)_\text{rad} > \left(\frac{\text{d}\ln T}{\text{d}\ln P}\right)_\text{ad} + \left(\frac{\partial\ln T}{\partial\ln\mu}\right)_P \frac{\text{d}\ln\mu}{\text{d}\ln P},
\label{eq:led}
\end{equation}
where $\text{d}\ln\mu/\text{d}\ln P$ stands for the gradient of mean molecular weight $\mu$ caused by inhomogeneous chemical composition in the stellar interior. 

It is known that layers satisfying the Schwarzschild condition (eq.~\ref{eq:sch}) but not the Ledoux criterion (eq.~\ref{eq:led}) are thermally unstable \citep[overstable; see][]{Kato1966}. This is one of the reasons why the Schwarzschild criterion is preferred in the evolution calculations. It is not clear, however, how efficiently the vibrational thermal instability should mix layers.

Convection in the Geneva stellar evolution code is treated as follows: the limits of convective regions are defined using either the Ledoux or the Schwarzschild criterion. During the main sequence and core helium-burning phase, the convective core is extended by a fraction of $0.1H_P$, where $H_P$ is the pressure scale height, evaluated at the strict Ledoux or Schwarzschild limit respectively. In this extension of the convective core, the thermal gradient is assumed to be the adiabatic one. Chemical  mixing inside convective zone is assumed to be very efficient and fast compared to the nuclear timescale. The chemical composition is thus homogenised inside convective regions, at least during the first stages of nuclear burning (as is the case for all the models computed in this work).

In this paper, we use different sets of models computed with the Schwarzschild or with the Ledoux criterion. In case the Ledoux one is used, we assume that there is no semi-convection in the regions where the matter is Schwarzschild unstable but Ledoux stable (note however that in case we deal with a model including the effects of rotation, there is a slow mixing of chemical elements inside these regions due to rotational mixing). There are poor constraints on the efficiency of semi-convection in stars, so that we make the decision to compute only extreme cases : pure Ledoux models without semi-convection, mimicking models where semi-convection is very inefficient, and pure Schwarzschild models, corresponding to Ledoux models with infinitely efficient semi-convection. Models with intermediate semi-convective efficiency would fall somewhere in between these two cases \citep[a discussion on the effect of varying the efficiency of semi-convection in massive star models can be found in][however in the framework of the MESA code]{Kaiser2020a}. Let us also emphasise here that the post-MS evolution of massive star is extremely dependent on the detailed mixing scheme adopted in stellar evolution codes, which makes difficult to converge towards firm conclusions \citep[e.g.][]{Higgins2019a,Schootemeijer2019a}.

The stability of radial pulsations is computed in the same way as in \citet{Saio2013}. Although the effects of rotation are taken into account in the evolution models, no mechanical effects of rotation are included in calculating the stability of radial pulsations. This is justified because, as discussed in the Appendix in \citet{Saio2013}, the rotation  speed is always very slow in the envelopes of supergiant stars, to which the amplitude of radial pulsation is confined.

\section{Evolutionary models with Schwarzschild and Ledoux criteria for $Z=0.014$}

\begin{figure}
\includegraphics[width=0.49\textwidth]{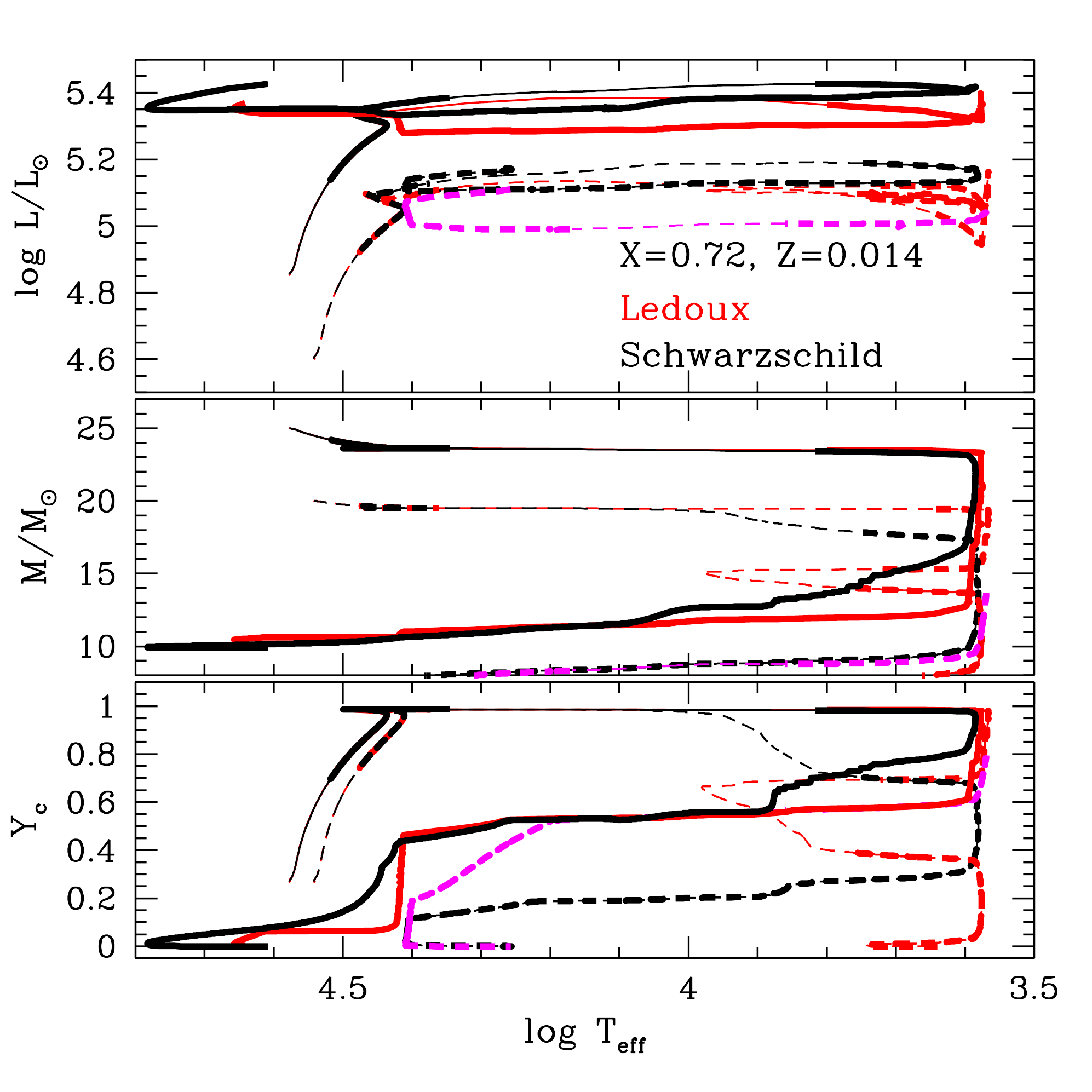}
\caption{Evolutionary tracks of $20$ (dashed lines) and $25\,\text{M}_\odot$ (solid lines) models until central helium exhaustion computed with the Ledoux (red lines) and the Schwarzschild (black lines) criteria for the occurrence of convection for an initial composition of $(X,Z) = (0.72,0.014)$. The magenta line corresponds to the Ledoux $20\,M_\sun$ model computed with an enhanced mass-loss rate (2x our standard rates). Top panel is the HR diagram, the middle and the bottom panels show changes in mass and in the central helium abundance, $Y_\text{c}$, respectively, as a function of $T_\text{eff}$ Along the thick-line parts radial pulsations are excited.
}
\label{fig:te_Lmyc}
\end{figure}

 \begin{figure*}[!h]
\includegraphics[width=0.49\textwidth]{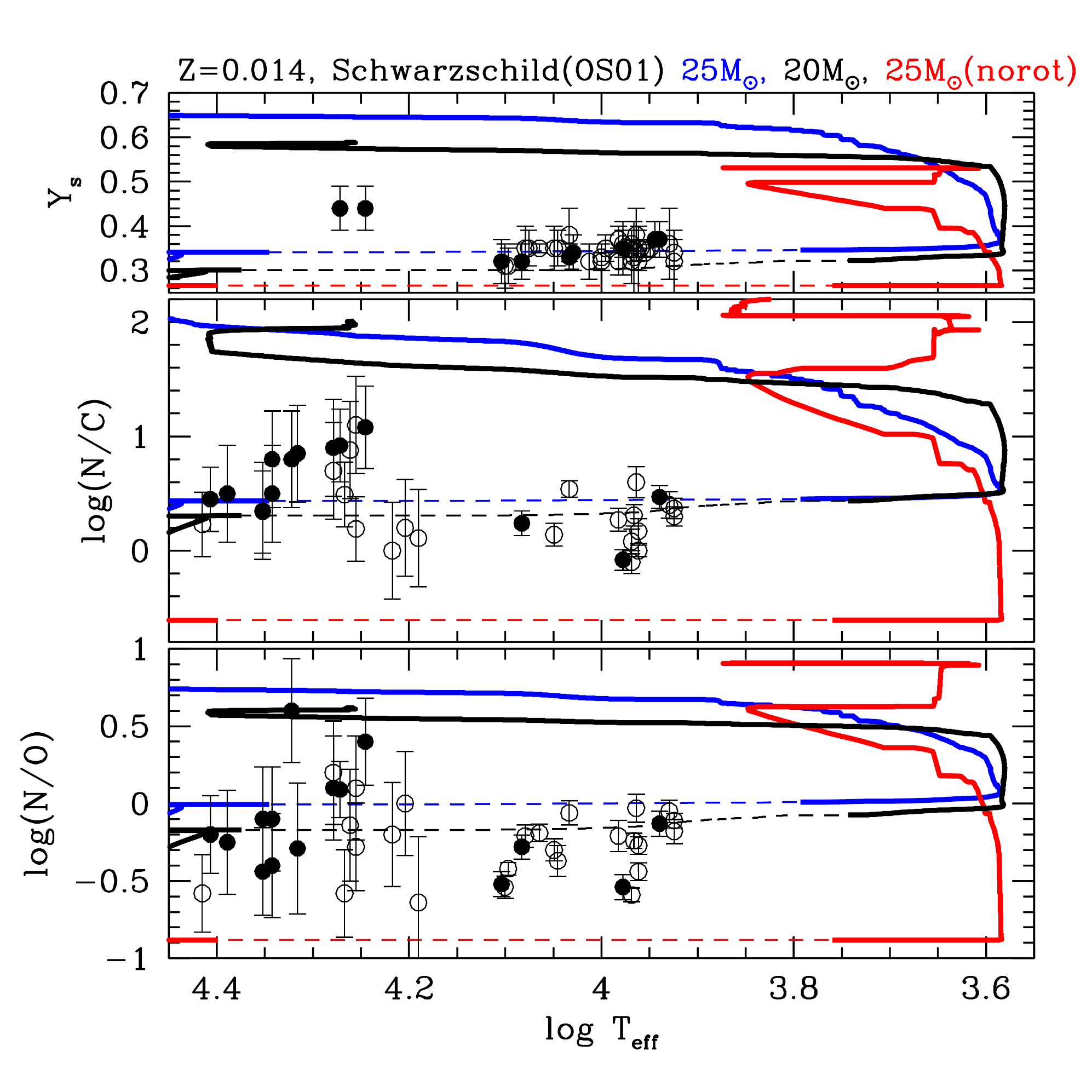}
\includegraphics[width=0.49\textwidth]{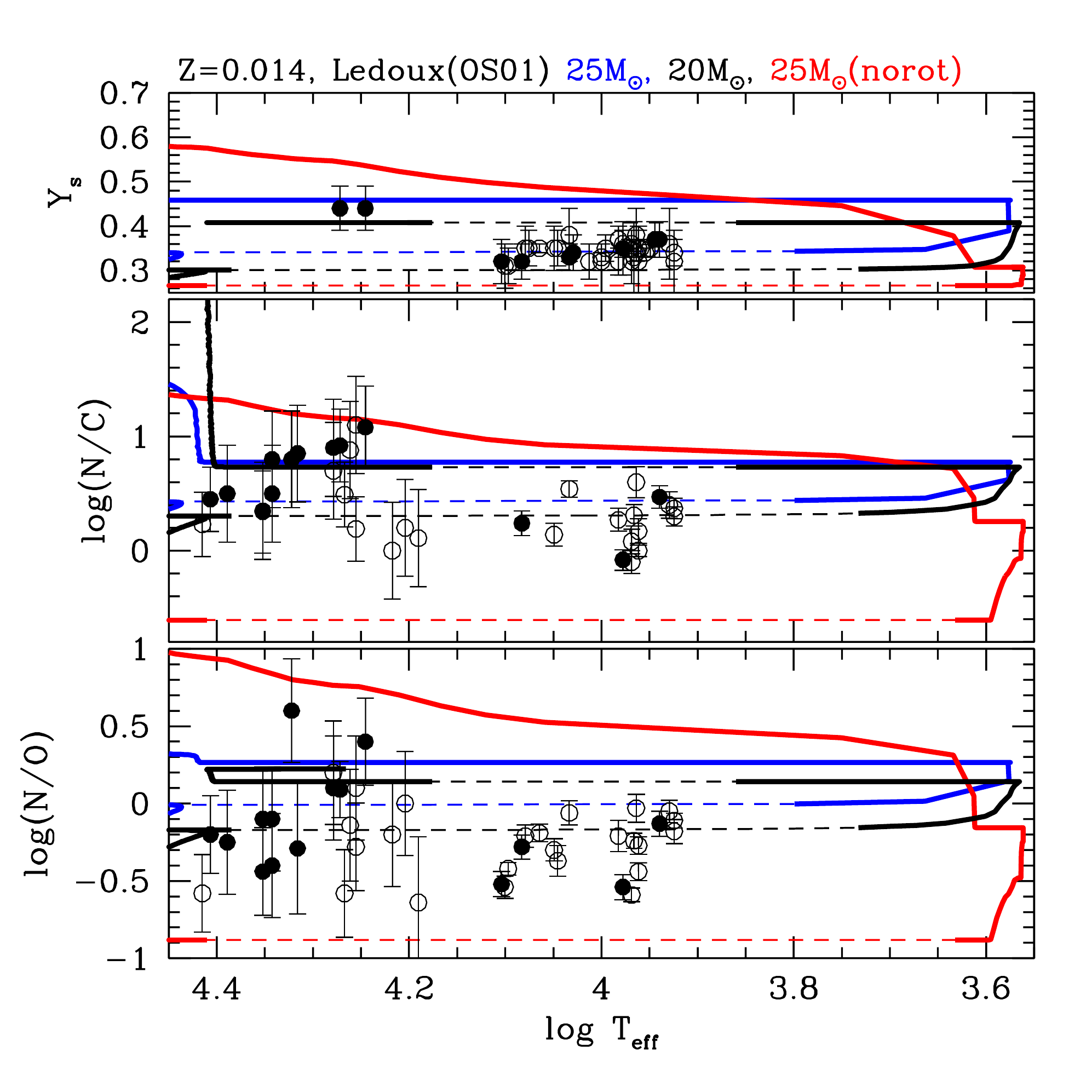}
\caption{Surface CNO and helium abundances predicted from $20\,\text{M}_\sun$ (with rotation, black lines) and $25\,\text{M}_\sun$ (with rotation in blue, without rotation in red line) evolutionary models are compared with observed surface CNO and helium abundances of Galactic blue supergiants, where $\text{N}/\text{C}$ and $\text{N}/\text{O}$ are number ratios, and $Y_\text{s}$ is mass fraction of helium at the surface. Solid parts of the lines indicate where radial pulsations are excited. Filled circles represent $\alpha$ Cyg variables, while open circles are for non-variable BSGs. Left panels are for models with the Schwarzschild condition, while right panels are for models with the Ledoux condition.
}
\label{fig:CNOys}
\end{figure*}

Fig.~\ref{fig:te_Lmyc} shows evolutions of global parameters until the core-helium exhaustion for $20$ and $25\,\text{M}_\odot$ models of $Z=0.014$ (corresponding to a solar metallicity) computed by using the Ledoux (red lines) and Schwarzschild (black lines) criteria. The Geneva evolution code was used as in \citet{Saio2013} and \citet{Georgy2014}. Wind mass loss is considered in the same way as described in our previous works \citep[see also][for more details]{Ekstrom2012}. In addition to these standard models, and for comparison purpose, we computed three additional models : a $20\,M_\sun$ computed with the Ledoux criterion and a mass-loss rate multiplied by $2$ with respect to our usual mass-loss rates with an initial rotation rate on the zero-age main-sequence (ZAMS) equal to $40\%$ of the critical velocity, and two models of $25\,M_\sun$ computed with the Ledoux criterion and with an overshoot increased to $0.3H_P$ (non-rotating and with an initial rotation rate on the ZAMS equal to $40\%$ of the critical velocity).

During the main-sequence evolution, the models with either criteria are nearly identical. This is caused by the following : a) in our models, the chemical composition is initially homogeneous, so that the chemical gradient $\nabla_\mu$ is null everywhere; b) it implies that in the initial models, both criteria lead to the same convective boundary; c) during the MS, the convective core is receding. As the chemical composition inside convective zones is homogeneous, it means that in the convective core, Schwarzschild and Ledoux criteria coincide, making the models computed with either criterion identical.

During the post main-sequence evolution, the models using the Ledoux criterion are slightly less luminous, but overall evolutions of global parameters are similar irrespective to the convection criteria (note however that the crossing-time of the Hertzsprung-Russell diagram (HRD) can be different, due to the activity of the intermediate H-burning convective zone). After the main-sequence, they evolve to the red supergiant region, where they burn a significant fraction of helium in the core and, at the same time, lose considerable amount of mass. When the helium core occupies about $70\%$ of the total mass, the stars will evolve bluewards, crossing the HRD for the second time \citep{Giannone1967a,Farrell2020a}. They will thus have another BSG stage, where they consumes the remaining helium in the core. Let us mention here that the Ledoux $20\,M_\sun$ model at solar metallicity computed with standard mass-loss rates does not lose enough mass during the RSG phase to evolve back to the BSG region of the HRD. As we are interested in stars evolving back to the blue after a RSG phase, we will consider in what follows the Ledoux $20\,M_\sun$ with enhanced mass loss: this model is crossing the HRD after its RSG stage and reaches regions in the HRD where the effective temperature corresponds to BSG stars. The Ledoux $20\,M_\sun$ model computed with the standard mass-loss rates makes instead a small blue loop and ends its evolution in cooler regions (see red dashed lines in Fig.~\ref{fig:te_Lmyc}). All the other models discussed in this work are computed assuming the standard mass-loss rates, as in \citet{Ekstrom2012}.

\longtab[1]{
\begin{longtable}{lrcccccccc}
\caption{Adopted parameters and variability types for Galactic blue supergiants}\label{tab:sum} \\
\hline
Name & $T_\text{eff}\,\text{(K)}$ & $\log\left(L/L_\odot\right)$ & $\log{g}_\text{F}$ & var & Ys   & $\epsilon$(C)   & $\epsilon$(N)  & $\epsilon$(O)  & Refs \\
\hline
\endfirsthead
\caption{Continued.}\\
\hline
Name & $T_\text{eff}\,\text{(K)}$ & $\log\left(L/L_\odot\right)$ & $\log{g}_\text{F}$ & var & Ys   & $\epsilon$(C)   & $\epsilon$(N)  & $\epsilon$(O)  & Refs \\
\hline
\endhead
\hline
\endfoot
HD 210221  &  8400$\pm$150  & 4.593$\pm$0.063 & 1.70$\pm$0.10 &  0 &   0.34$\pm$0.03 &   8.22$\pm$0.06 &   8.52$\pm$0.06 &   8.70$\pm$0.05 & 1\\
HD 213470  &  8400$\pm$150  & 5.055$\pm$0.108 & 1.60$\pm$0.10 &  0 &   0.32$\pm$0.07 &   8.17$\pm$0.08 &   8.54$\pm$0.04 &   8.65$\pm$0.03 & 1\\
HD 13476   &  8500$\pm$150  & 4.936$\pm$0.096 & 1.68$\pm$0.10 &  0 &   0.36$\pm$0.08 &   8.18$\pm$0.11 &   8.58$\pm$0.04 &   8.63$\pm$0.06 & 1\\
HD 197345  &  8700$\pm$150  & 4.742$\pm$0.124 & 1.44$\pm$0.10 &  1 &   0.37$\pm$0.04 &   8.09$\pm$0.07 &   8.56$\pm$0.07 &   8.69$\pm$0.04 & 1\\ 
HD 207260  &  8800$\pm$150  & 4.510$\pm$0.188 & 1.57$\pm$0.10 &  0 &   0.37$\pm$0.04 &   8.22$\pm$0.08 &     -       &     -       & 2\\
HD 102878  &  8900$\pm$150  & 4.580$\pm$0.134 & 1.70$\pm$0.10 &  2 &   0.35$\pm$0.02 &   8.26$\pm$0.12 &     -       &     -       & 2,a\\
HD 165784  &  9000$\pm$200  & 4.853$\pm$0.095 & 1.68$\pm$0.11 &  0 &   0.34$\pm$0.01 &   8.39$\pm$0.05 &     -       &     -       & 2\\
HD 91533   &  9100$\pm$150  & 5.171$\pm$0.133 & 1.66$\pm$0.10 &  0 &   0.35$\pm$0.02 &   8.20$\pm$0.03 &     -       &     -       & 2\\
HD 14433   &  9150$\pm$150  & 4.969$\pm$0.099 & 1.55$\pm$0.10 &  0 &   0.32$\pm$0.07 &   8.23$\pm$0.04 &   8.23$\pm$0.03 &   8.67$\pm$0.05 & 1\\
HD 111613  &  9150$\pm$150  & 5.218$\pm$0.178 & 1.60$\pm$0.10 &  0 &   0.35$\pm$0.05 &   8.29$\pm$0.10 &   8.46$\pm$0.04 &   8.73$\pm$0.04 & 1\\
HD 12953   &  9200$\pm$200  & 5.460$\pm$0.214 & 1.29$\pm$0.11 &  0 &   0.34$\pm$0.01 &     -       &   8.45$\pm$0.03 &     -       & 2\\
HD 195324  &  9200$\pm$150  & 4.203$\pm$0.045 & 1.99$\pm$0.10 &  0 &   0.38$\pm$0.06 &   8.10$\pm$0.11 &   8.70$\pm$0.08 &   8.73$\pm$0.04 & 1\\
HD 207673  &  9250$\pm$100  & 4.468$\pm$0.051 & 1.93$\pm$0.10 &  0 &   0.33$\pm$0.08 &   8.17$\pm$0.09 &   8.48$\pm$0.03 &   8.72$\pm$0.04 & 1\\
HD 80057   &  9300$\pm$150  & 4.754$\pm$0.100 & 1.88$\pm$0.10 &  0 &   0.36$\pm$0.03 &   8.24$\pm$0.10 &   8.32$\pm$0.04 &     -       & 2\\
HD 187983  &  9300$\pm$250  & 5.198$\pm$0.163 & 1.73$\pm$0.16 &  0 &   0.32$\pm$0.05 &   8.29$\pm$0.09 &   8.19$\pm$0.04 &   8.78$\pm$0.02 & 1\\
HD 14489   &  9350$\pm$250  & 4.595$\pm$0.133 & 1.57$\pm$0.16 &  0 &   0.35$\pm$0.02 &     -       &   8.53$\pm$0.06 &     -       & 2\\
HD 13744   &  9500$\pm$250  & 4.691$\pm$0.089 & 1.64$\pm$0.14 &  0 &   0.36$\pm$0.03 &     -       &     -       &     -       & 2\\
HD 92207   &  9500$\pm$200  & 5.106$\pm$0.146 & 1.29$\pm$0.11 &  1 &   0.35$\pm$0.06 &   8.33$\pm$0.08 &   8.25$\pm$0.04 &   8.79$\pm$0.07 & 1,b,c\\
HD 87737   &  9600$\pm$150  & 3.530$\pm$0.196 & 2.12$\pm$0.10 &  0 &   0.37$\pm$0.03 &   8.25$\pm$0.06 &   8.52$\pm$0.08 &   8.73$\pm$0.06 & 1\\
HD 166167  &  9600$\pm$150  & 4.032$\pm$0.164 & 2.07$\pm$0.10 &  0 &   0.32$\pm$0.03 &     -       &     -       &     -       & 2\\
HD 149077  &  9900$\pm$150  & 4.066$\pm$0.075 & 2.22$\pm$0.10 &  0 &   0.35$\pm$0.03 &     -       &   8.45$\pm$0.05 &     -       & 2\\
HD 20041   & 10000$\pm$200  & 4.740$\pm$0.093 & 1.65$\pm$0.11 &  0 &   0.32$\pm$0.02 &     -       &   8.30$\pm$0.04 &     -       & 2\\
HD 46300   & 10000$\pm$200  & 3.440$\pm$0.140 & 2.15$\pm$0.11 &  0 &   0.33$\pm$0.03 &     -       &   8.43$\pm$0.07 &     -       & 2\\
HD 39970   & 10300$\pm$200  & 4.743$\pm$0.082 & 1.65$\pm$0.11 &  0 &   0.32$\pm$0.04 &     -       &   8.15$\pm$0.10 &     -       & 2\\
HD 223960  & 10700$\pm$200  & 5.498$\pm$0.119 & 1.48$\pm$0.10 &  2 &   0.34$\pm$0.04 &     -       &   8.55$\pm$0.07 &     -       & 2,a,c\\
HD 21291   & 10800$\pm$200  & 4.691$\pm$0.162 & 1.52$\pm$0.10 &  1 &   0.33$\pm$0.03 &     -       &   8.46$\pm$0.04 &     -       & 2,d\\
HD 202850  & 10800$\pm$200  & 3.967$\pm$0.123 & 1.72$\pm$0.10 &  2 &   0.38$\pm$0.06 &   8.16$\pm$0.04 &   8.70$\pm$0.06 &   8.76$\pm$0.05 & 1,a\\
HD 149076  & 11100$\pm$200  & 4.543$\pm$0.100 & 1.87$\pm$0.10 &  0 &   0.35$\pm$0.04 &     -       &   8.43$\pm$0.09 &   8.80$\pm$0.04 & 2\\
HD 212593  & 11200$\pm$200  & 3.959$\pm$0.137 & 1.90$\pm$0.10 &  0 &   0.35$\pm$0.05 &   8.30$\pm$0.08 &   8.44$\pm$0.06 &   8.74$\pm$0.04 & 1\\
HD 106068  & 11600$\pm$200  & 4.750$\pm$0.069 & 1.64$\pm$0.10 &  0 &   0.35$\pm$0.01 &     -       &   8.60$\pm$0.04 &   8.79$\pm$0.04 & 2\\
BD+602582 & 11900$\pm$200  & 4.903$\pm$0.140 & 1.55$\pm$0.10 &  0 &   0.35$\pm$0.04 &     -       &   8.53$\pm$0.05 &     -       & 2\\
HD 105071  & 12000$\pm$150  & 4.805$\pm$0.060 & 1.53$\pm$0.10 &  2 &   0.35$\pm$0.05 &     -       &   8.55$\pm$0.06 &   8.76$\pm$0.04 & 2,b\\
HD 34085   & 12100$\pm$150  & 5.062$\pm$0.079 & 1.42$\pm$0.10 &  1 &   0.32$\pm$0.04 &   8.23$\pm$0.09 &   8.47$\pm$0.06 &   8.75$\pm$0.05 & 1,b\\
HD 186745  & 12500$\pm$200  & 5.144$\pm$0.068 & 1.41$\pm$0.10 &  0 &   0.31$\pm$0.04 &     -       &   8.35$\pm$0.04 &   8.77$\pm$0.01 & 2\\
HD 12301   & 12600$\pm$200  & 4.587$\pm$0.067 & 1.75$\pm$0.10 &  0 &   0.31$\pm$0.05 &     -       &   8.14$\pm$0.04 &   8.68$\pm$0.06 & 2\\
HD 208501  & 12700$\pm$200  & 4.790$\pm$0.067 & 1.43$\pm$0.10 &  1 &   0.32$\pm$0.05 &     -       &   8.24$\pm$0.08 &   8.76$\pm$0.01 & 2,e\\
HD 164353  & 15500$\pm$1000 & 5.213$\pm$0.258 & 1.99$\pm$0.27 &  0 &     -       &   7.78$\pm$0.30 &   7.89$\pm$0.30 &   8.53$\pm$0.30 &4\\
HD 14134   & 16000$\pm$1000 & 5.305$\pm$0.092 & 1.23$\pm$0.18 &  2 &     -       &   8.25$\pm$0.30 &   8.45$\pm$0.30 &   8.45$\pm$0.15 &3,b\\
HD 198478  & 16500$\pm$500  & 5.166$\pm$0.151 & 1.28$\pm$0.16 &  2 &     -       &   8.25$\pm$0.30 &   8.25$\pm$0.30 &   8.45$\pm$0.15 &3,b\\
HD 152236  & 17600$\pm$500  & 5.714$\pm$0.307 & 1.12$\pm$0.16 &  1 &   0.44$\pm$0.05 &   7.69$\pm$0.30 &   8.77$\pm$0.20 &   8.37$\pm$0.20 &3,5,g \\
HD 14143   & 18000$\pm$1000 & 5.431$\pm$0.076 & 1.23$\pm$0.18 &  2 &     -       &   7.60$\pm$0.30 &   8.70$\pm$0.30 &   8.60$\pm$0.15 &3,b\\
HD 206165  & 18000$\pm$500  & 5.094$\pm$0.192 & 1.23$\pm$0.25 &  2 &     -       &   7.96$\pm$0.20 &   8.15$\pm$0.20 &   8.43$\pm$0.20 &4,b\\
HD 14818   & 18250$\pm$500  & 5.824$\pm$0.151 & 1.36$\pm$0.11 &  2 &     -       &   7.66$\pm$0.30 &   8.54$\pm$0.30 &   8.68$\pm$0.20 &3,4,b\\
HD 193183  & 18500$\pm$1000 & 5.323$\pm$0.096 & 1.56$\pm$0.27 &  0 &     -       &   7.66$\pm$0.20 &   8.15$\pm$0.20 &   8.73$\pm$0.20 &4\\
HD 190603  & 18700$\pm$1000 & 6.111$\pm$0.251 & 1.15$\pm$0.18 &  1 &   0.44$\pm$0.05 &   7.84$\pm$0.30 &   8.76$\pm$0.10 &   8.67$\pm$0.15 &3,4,5,b,e\\
HD 41117   & 19000$\pm$1000 & 4.781$\pm$0.135 & 1.23$\pm$0.18 &  1 &     -       &   7.65$\pm$0.30 &   8.55$\pm$0.30 &   8.45$\pm$0.15 &3,b\\
HD 194279  & 19000$\pm$1000 & 5.699$\pm$0.037 & 1.18$\pm$0.18 &  0 &     -       &   7.95$\pm$0.30 &   8.65$\pm$0.30 &   8.45$\pm$0.15 &3\\
HD 13854   & 20700$\pm$2000 & 5.805$\pm$0.164 & 1.24$\pm$0.30 &  1 &     -       &   7.66$\pm$0.30 &   8.51$\pm$0.30 &   8.80$\pm$0.30 &3,b\\
HD 14956   & 21000$\pm$1000 & 6.088$\pm$0.164 & 1.21$\pm$0.17 &  1 &     -       &   7.95$\pm$0.30 &   8.75$\pm$0.30 &   8.15$\pm$0.15 &3,f\\
HD 91316   & 22000$\pm$1000 & 4.659$\pm$0.215 & 1.18$\pm$0.17 &  1 &     -       &   7.50$\pm$0.30 &   8.30$\pm$0.30 &   8.40$\pm$0.15 &3,b\\
HD 148688  & 22000$\pm$1000 & 5.536$\pm$0.258 & 1.23$\pm$0.17 &  1 &     -       &   7.65$\pm$0.30 &   8.15$\pm$0.30 &   8.55$\pm$0.15 &3,b\\
HD 2905    & 22500$\pm$1000 & 4.881$\pm$0.144 & 1.29$\pm$0.17 &  1 &     -       &   7.82$\pm$0.30 &   8.16$\pm$0.20 &   8.60$\pm$0.20 &3,4,b,e\\
HD 154090  & 22500$\pm$500  & 5.285$\pm$0.165 & 1.24$\pm$0.15 &  2 &     -       &   7.95$\pm$0.30 &   8.30$\pm$0.30 &   8.40$\pm$0.15 &3,b\\
HD 91943   & 24500$\pm$500  & 5.666$\pm$0.412 & 1.24$\pm$0.20 &  1 &     -       &   7.65$\pm$0.30 &   8.15$\pm$0.30 &   8.40$\pm$0.15 &3,b\\
HD 115842  & 25500$\pm$500  & 5.608$\pm$0.090 & 1.22$\pm$0.15 &  1 &     -       &   7.80$\pm$0.20 &   8.25$\pm$0.20 &   8.45$\pm$0.15 &3,b\\
HD 185859  & 26000$\pm$1000 & 5.033$\pm$0.045 & 1.47$\pm$0.26 &  0 &     -       &   7.72$\pm$0.20 &   7.95$\pm$0.20 &   8.53$\pm$0.15 &4\\
\hline

\end{longtable}
\tablebib{
 (1) \citet{Przy2010}; (2) \citet{Firnstein2012};  (3) \citet{Crowther2006};
 (4) \citet{Searle2008}; (5) \citet{Clark2012};
 (a) \citet{Koen2002};   (b) \citet{Lefevre2009}; (c) \citet{Rimoldini2012};
 (d) \citet{Dubath2011}; (e) \citet{Percy1983}; (f) \citet{Burki1978}; 
 (g) \citet{Sterken1997}
}
\tablefoottext{$^*$}{References of (1)--(5) are for physical parameters and surface compositions, and (a)--(g) for the variability types.}
}

\section{Surface abundances of Galactic blue supergiants}

\subsection{Models with standard parameters}

\begin{figure*}[!h]
\includegraphics[width=.98\textwidth]{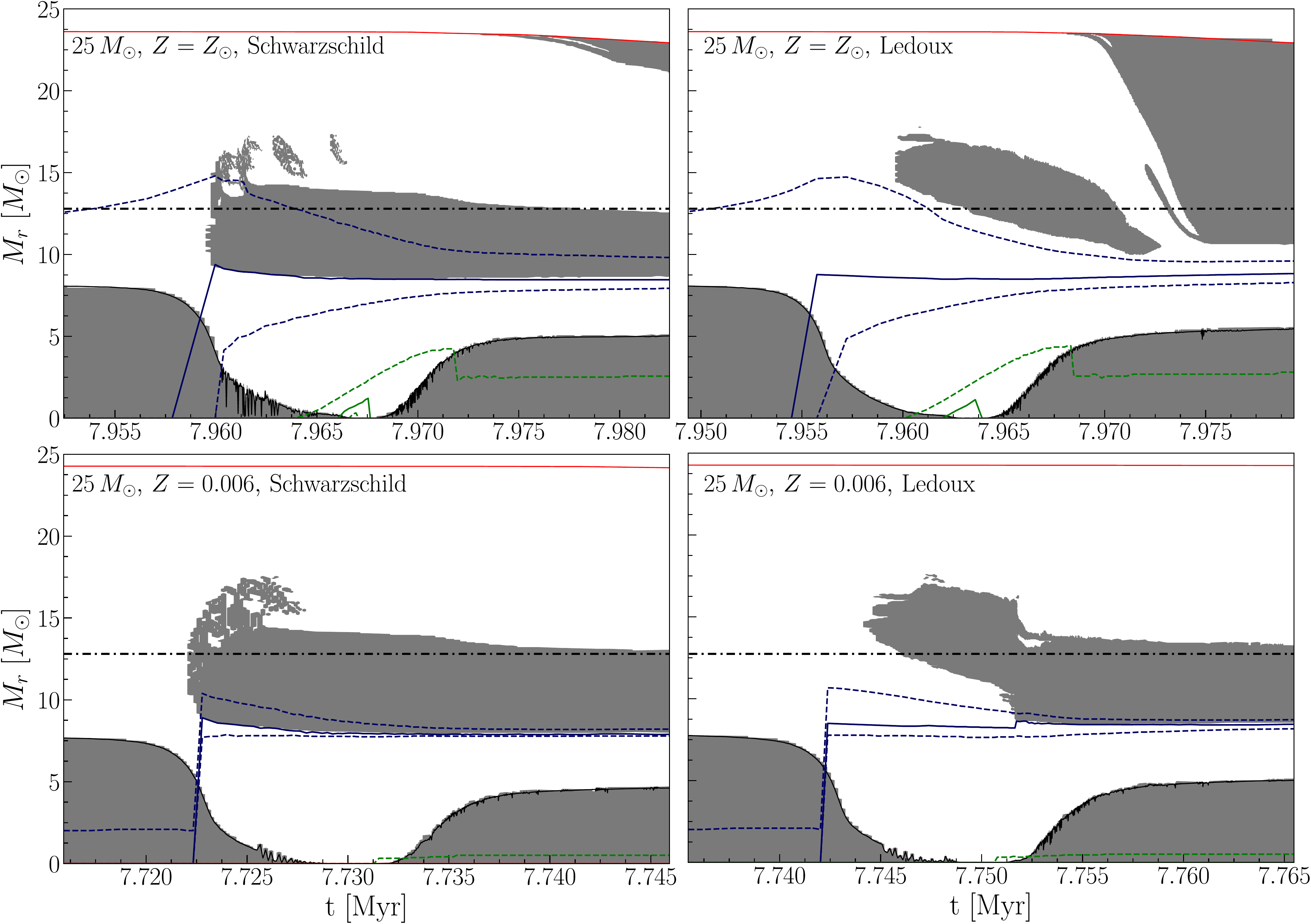}
\caption{Kippenhahn diagrams for four representative models of $25\,M_\sun$ : solar metallicity (top row), and $Z=0.006$ (bottom row); computed with the Schwarzschild criterion (left column) or Ledoux criterion (right column). Grey zones indicate that this part of the star is convective. The mass coordinate of the surface is shown with a red line. Maxima of energy generation rate are indicated for H burning (blue solid line) and He burning (green solid line). The dashed lines indicate, for H and He-burning respectively, the level where the energy generation rate reaches $100\,[\text{erg}\,\text{s}^{-1}\,\text{g}^{-1}]$. The dot-dashed line show the mass-coordinate of the convective core on the ZAMS. It roughly shows the upper boundary of the region in which H-burning has proceed, and therefore has increased the chemical gradient (due to rotational mixing, the exact extension of the region where a significant chemical gradient is present is slightly larger).
}
\label{fig:KippenAll}
\end{figure*}

Fig.~\ref{fig:CNOys} compares surface N/C and N/O (number) ratios and helium abundances, $Y_\text{s}$ (mass fraction) of Galactic blue supergiants with theoretical models with rotation ($25\,M_\sun$, blue line and $20\,M_\sun$, black line) and without rotation ($25\,M_\sun$, red line). The left panels are for models with the Schwarzschild criterion for convection, while  the right panels are for the models with  the Ledoux criterion. Radial pulsations are excited in  the solid-line parts. Physical parameters of observed BSGs shown here can be found in Table~\ref{tab:sum} at the end of the article.

The surface helium and CNO abundances in the models after significant mass loss in the red supergiant stage depend strongly on the adopted criterion for convection. As discussed in \citet{Georgy2014} the difference comes from the location of the intermediate convective shell located just above the hydrogen burning shell during the blue supergiant stage evolving toward the red supergiant region (group 1):

\paragraph{Schwarzschild criterion :} At the end of the MS, the convective hydrogen-burning core recedes and disappears. Hydrogen burning jumps in a shell located higher inside the star (see the top-left panel of Fig.~\ref{fig:KippenAll}). It creates an associated convective shell which is active enough to sustain the star and prevent a rapid crossing of the HRD. This convective shell is able to transport matter processed by CNO cycle to a high level inside the star. When the star crosses the HRD for a second time towards the BSG region, it has lost a lot of mass during the RSG phase, uncovering layers highly processed by CNO cycle, i.e. with high $\text{N}/\text{C}$ and $\text{N}/\text{O}$ ratios.

\begin{figure}
\includegraphics[width=.48\textwidth]{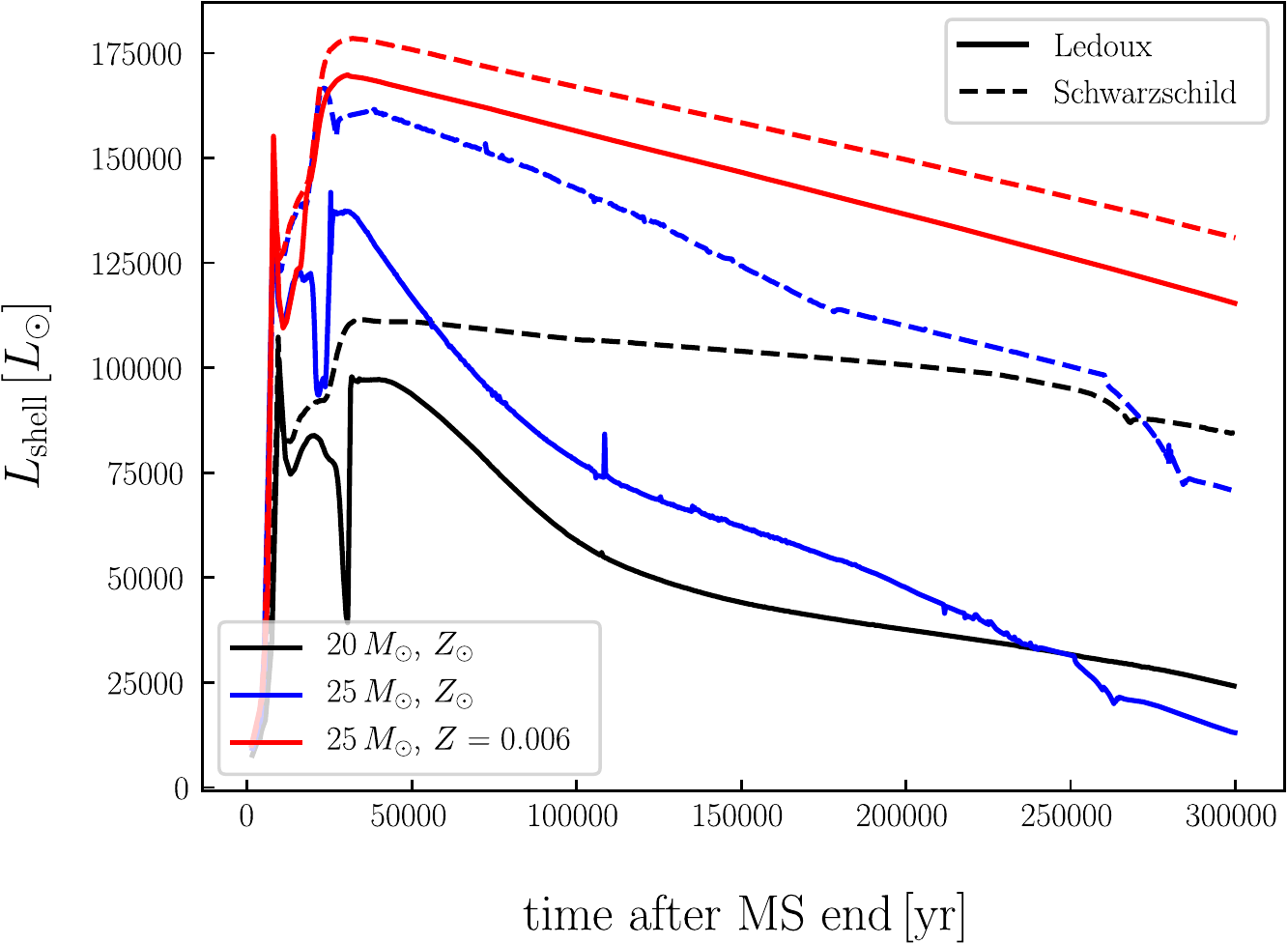}
\caption{Time evolution of the luminosity associated with the hydrogen-burning shell during the first couple of $10^5$ years after the end of the MS. Solid lines are for models computed with the Ledoux criterion, and dashed lines are for models computed with the Schwarzschild criterion. Black lines are for $20\,M_\sun$ models at solar metallicity, blue lines for $25\,M_\sun$ models at solar metallicity, and red lines for  $25\,M_\sun$ models at a metallicity $Z=0.006$.
}
\label{fig:LumShell}
\end{figure}

\paragraph{Ledoux criterion :} As in the case of the Schwarzschild criterion, the H-burning convective core disappears at the end of the MS, and the maximum of H-burning migrates at a higher level inside the star. However, it occurs in a region of the star that was previously occupied by the receding H-burning core, and therefore with a non-zero chemical gradient (see the position of the dot-dashed line, which shows the extension of the convective core on the ZAMS : due to H-burning, a chemical gradient is progressively built during the MS below this line). This gradient prevents the appearance of a convective zone at the same depth as with the Schwarzschild criterion. Convection is only able to develop in layers closer to the surface (see the top-right panel of Fig.~\ref{fig:KippenAll}), from where it slowly erodes the chemical gradient, making its lower boundary moving towards the centre of the star. It makes the hydrogen-burning shell much less active compared to the Schwarzschild case, as illustrated in Fig.~\ref{fig:LumShell}. The luminosity of the shell is not sufficient to sustain the star, which crosses the HRD on a very short timescale : the outer layers expand, a convective envelope appears from the surface and develops in depth, while the intermediate convective shell fades away. The matter inside the intermediate convective shell is less processed compared to the Schwarzschild case, making the $\text{N}/\text{C}$ and $\text{N}/\text{O}$ ratios remain smaller at the time when these layers are exposed at the surface because of the mass loss (see Fig.~\ref{fig:CNOys}). This difference is particularly noticeable in the mass range around $20$ to $25\,M_\sun$ (higher mass models have a tendency to avoid the red supergiant phase remaining in the blue part of the HRD during the whole evolution, and lower mass models quickly cross the HRD, but don't lose enough mass during the RSG phase to evolve back to the blue later on).

For the late B and A type supergiants ($4.15 \ga \log T_\text{eff} \ga 3.9$) the surface helium abundances and $\text{N}/\text{C}$ and $\text{N}/\text{O}$ ratios are, irrespective to their variability, close to the values of models of the group 1. This is at odds with our explanation of  the $\alpha$ Cyg variables as blue supergiants after the red supergiant stages \citep{Saio2013}, in which $\alpha$ Cyg variables should have, in particular for Schwarzschild criterion, higher $Y_\text{s}$, and $\text{N}/\text{C}$ and $\text{N}/\text{O}$ ratios. However, for the models with the Ledoux criterion, the discrepancy is considerably reduced since the differences of these quantities for the models before and after the red supergiant stage are much smaller than for the models with the Schwarzschild criterion \citep{Georgy2014}. Due to the much longer time spent on the second crossing compared to the first one (see Table~\ref{TableDuration}), our computations show that there is also a much greater probability to observe a star of the BSG2 group, since the duration of the BSG2 phase is at least 30 times longer than the BSG1 one. The comparison between the tracks and the observation should thus focus on the second crossing.

\begin{table*}
\begin{center}
\caption{Duration of the group 1 BSG\tablefootmark{a} phase and of the group 2 BSG\tablefootmark{a} phase for different models computed with the Ledoux criterion for convection. These models have a solar metallicity ($Z_\sun=0.014$)}.
\label{TableDuration}
\begin{tabular}{ccc|cc}
mass & rotation & comment & BSG1 duration & BSG2 duration\\
$M_\sun$ & $V/V_\text{crit, ini}$ & & kyr & kyr\\\hline
20 & 0.4 & increased $\dot{M}_\text{RSG}$\rule{0mm}{4mm} & $11.5$ & $500.3$\\
25 & 0.0 & - & $8.3$ & $306.0$\\
25 & 0.4 & - & $6.6$ & $375.1$\\
20 & 0.4 & overshoot $0.3H_P$ & $6.0$ & $442.4$
\end{tabular}
\end{center}\tablefoot{\tablefoottext{a}{Here we consider the model as a BSG if he has left the MS and if $3.9<\log(T_\text{eff})<4.4$.}}
\end{table*}

The surface helium abundance ($Y_\text{s}$) of Galactic early B supergiants ($\log T > 4.15$) are appreciably higher than later B and A supergiants, although only two cases are available; the surface $\text{N}/\text{C}$ and $\text{N}/\text{O}$ ratios of the early B $\alpha$ Cyg variables tend to be higher than those of non-variables, in contrast to the case of the late B and A type supergiants. These values of $Y_\text{s}$, $\text{N}/\text{C}$, and $\text{N}/\text{O}$ of the early type $\alpha$ Cyg variables agree with supergiant models returned from the red supergiant stage (group 2) obtained by using the Ledoux criterion, much better than those based on the Schwarzschild criterion. Models with the Schwarzschild criterion predict too high $Y_\text{s}$ and  $\text{N}/\text{C}$  values, in particular. Therefore, we conclude that the available spectroscopic surface abundances of Galactic blue supergiants in the literature indicate that the Ledoux criterion is better in reproducing the surface abundances of $\alpha$ Cyg variables, assuming that these stars are group 2 stars as deduced from their pulsational properties. However, despite our efforts so far, no model seems able to reproduce simultaneously the surface abundances of pulsating late type BSGs and their pulsational properties, even when considering models computed with the Ledoux criterion. Looking at Fig.~\ref{fig:CNOys}, we could wonder whether Schwarzschild models with lower initial mass could also fit the observed data. However, Schwarzschild models with masses between about $10\,M_\sun$ to $20\,M_\sun$ do not show bluewards evolution after the RSG phase \citep{Ekstrom2012}, unless very high mass-loss rate are considered \citep{Georgy2012a,Meynet2015a}.

\begin{figure}
\includegraphics[width=0.49\textwidth]{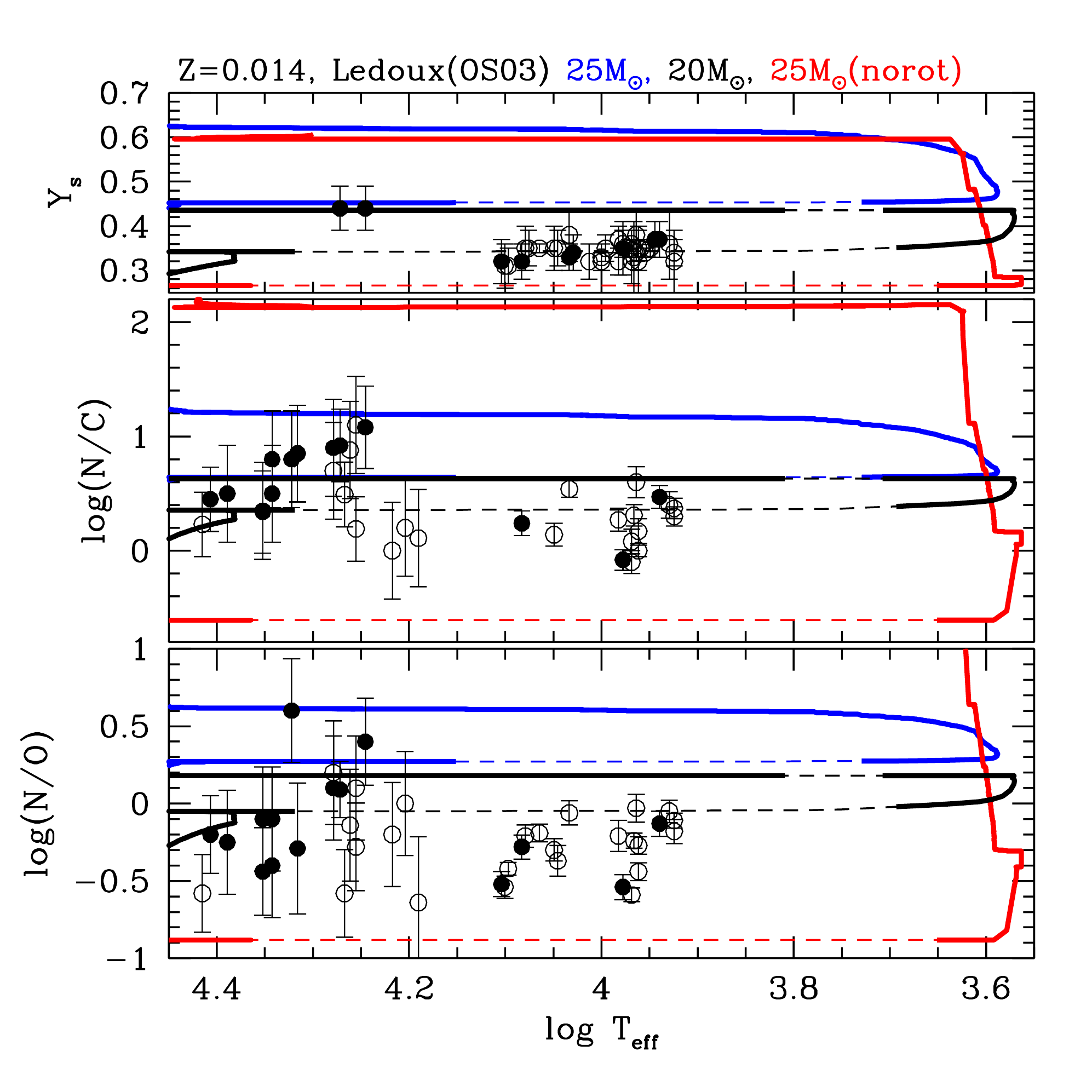}
\caption{The same as the right panel of Fig.~\ref{fig:CNOys} but for models with an increased overshoot of $0.3H_\text{P}$.}
\label{fig:CNOysOS03}
\end{figure}

In summary : our pulsation models for Galactic metallicity indicates that the only way of reproducing the observed pulsational properties of BSGs is to lose a large amount of mass during a previous RSG stage \citep{Saio2013}. These stars should therefore be members of group 2 BSGs. If this result is correct, then models computed with the Ledoux criterion for convection are in better agreement with the observed surface abundances, even if a perfect match is so far not reached. Also, our findings would indicate that stars with an initial mass of about $20\,M_\sun$ need a quite high mass-loss rate during the RSG stage to cross the Hertzsprung gap for a second time. This could be achieved in case the current mass-loss prescriptions for RSG are underestimated, or by binary interactions.

\subsection{Models with an increased overshoot}

The fraction of $H_P$ over which the convective core is extended is a free parameter of the code. The value of $0.1H_P$ generally used in this work is our standard value, and was calibrated by comparing the width of the MS obtained in our calculation with the observed width of MS around $2\,M_\sun$. It has been suggested that the overshooting as implemented in most stellar evolution code could be mass dependent and increase with the stellar mass \citep[e.g.][]{Castro2014a}. In this section, we explore the impact of increasing the fraction to $0.3H_P$ on the properties of BSGs, particularly on their pulsationnal properties and surface abundances.

On Fig~\ref{fig:CNOysOS03} are displayed evolutionnary tracks for models computed with the Ledoux criterion for convection ($25\,M_\sun$ without rotation in red, $25\,M_\sun$ with an initial rotation rate of $40\%$ of the critical one in blue, and $20\,M_\sun$ with an initial rotation rate of $40\%$ of the critical one in black). Comparing to the tracks with standard overshoot (see Fig.~\ref{fig:CNOys}), the agreement with the observations is slightly worse: particularly, the surface abundance of helium is systematically higher in the models, and the N/O and N/C ratios are also too high in the portions of the tracks where radial pulsations are excited (solid part of the tracks, to be compared with filled circles). The discussion about the duration of the BSG1 and BSG2 phases still holds in this case (see Table~\ref{TableDuration}).

Part of the explanation of this result resides in the bigger core during the MS produced by the increased overshoot. This reduces the distance between the edge of the convective core (where $Y$, $\text{N}/\text{C}$, and $\text{N}/\text{O}$ increase as evolution proceeds, due to the CNO cycle) and the surface. As a consequence, the chemical elements produced inside the convective core take a shorter time to reach the surface thanks to rotationnal mixing, making these quantities reaching higher value already during the first crossing of the HRD, and shifting up the overall tracks.

\section{Models of massive stars in the  LMC}
In the previous section, we found that the surface helium and CNO abundances of the Galactic blue supergiants agree with models based on the Ledoux criterion better than those with Schwarzschild criterion. In this section, we will discuss supergiant models with a LMC composition obtained with  the Ledoux and the Schwartzschild criteria.

\subsection{Evolution}
For the LMC star models, we have adopted a chemical composition of $(X,Z)=(0.738,0.006)$ (although we have also examined a model with $Z=0.008$, no qualitative changes occurred). Fig.~\ref{fig:te_Lmyc_LMC} compares evolution with the Schwarzschild criterion (left panel) and  the Ledoux criterion (right panel) for $20\,M_\sun$, $25\,M_\sun$, $30\,M_\sun$, and $35\,M_\sun$ models. These are rotating models with initial velocity of $0.4\,V_\text{c}$ (where $V_\text{c}$ means the critical rotation speed), except for the green track in right panel, which represents a non-rotating $20\,M_\sun$ model. Again, thick line parts indicate where radial pulsations are excited.

In contrast to the solar abundance case shown in Fig.~\ref{fig:te_Lmyc}, Fig.~\ref{fig:te_Lmyc_LMC} shows that in the case of  the LMC composition, core-helium burning tends to occur in a blue region of $4.2\ga\log T_\text{eff}\ga 4$, allowing for a significant amount of mass to be lost in this region of the HRD for models with $M \ge 25\,M_\sun$. Evolutions with the Ledoux criterion are similar to those with the Schwarzschild criterion when $M\le 25\,M_\odot$, while considerable differences occur in more massive cases. In the former case, after a considerable amount of central helium is consumed, the star evolves to the red-supergiant range where core-helium is exhausted.

In the case of the Schwarzschild criterion, $30$ and $35\,M_\sun$ models evolve to red or yellow supergiants after the main-sequence, where they start core-helium burning. After some amount of helium is consumed, they go back to the blue region to finish core-helium burning. In the case of the Ledoux criterion, on the other hand, they start core-helium burning at $\log T_\text{eff} \sim 4.2$ without becoming a red or yellow supergiant. The $30\,M_\sun$ model loses considerable mass during the helium burning and evolves to a red supergiant after most of the central helium is consumed, while the $35\,M_\sun$ model becomes hotter when it lost more than $10\,M_\odot$. 

The non-rotating model of $20\,M_\odot$ (green line) with the Ledoux criterion evolves very differently with a long blue loop extending as hot as $\log T_\text{eff} \approx 4.6$ after the red supergiant stage, which is rather similar to the cases of the solar metallicity shown in Fig.~\ref{fig:te_Lmyc}. The model ignites helium burning in the core and consumes significant helium as a red supergiant, during which it loses a mass of about $12\,M_\sun$. Then it becomes a blue supergiant and consumes remaining helium in the core. This property is however sensitive to the rotation rate; if we include a small initial rotation of $0.2\,V_\text{c}$, the loop shrinks significantly, extending only to $\log T_\text{eff}\approx 3.9$ (not shown in the figure for clarity purpose).
 
\subsection{Distribution of variable and non-variable supergiants of the LMC}

\begin{figure*}
\begin{center}
\includegraphics[width=0.49\textwidth]{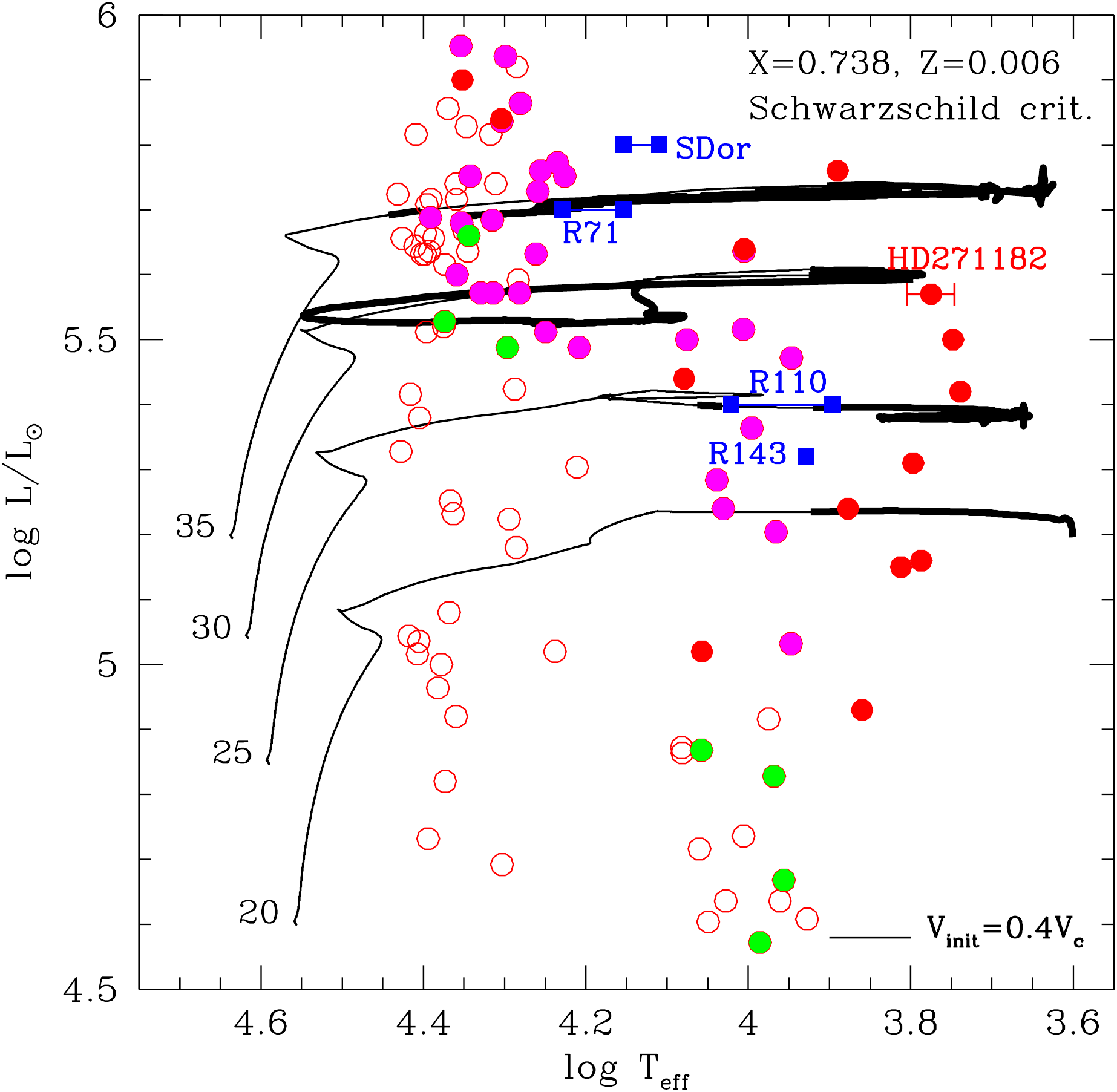}
\includegraphics[width=0.49\textwidth]{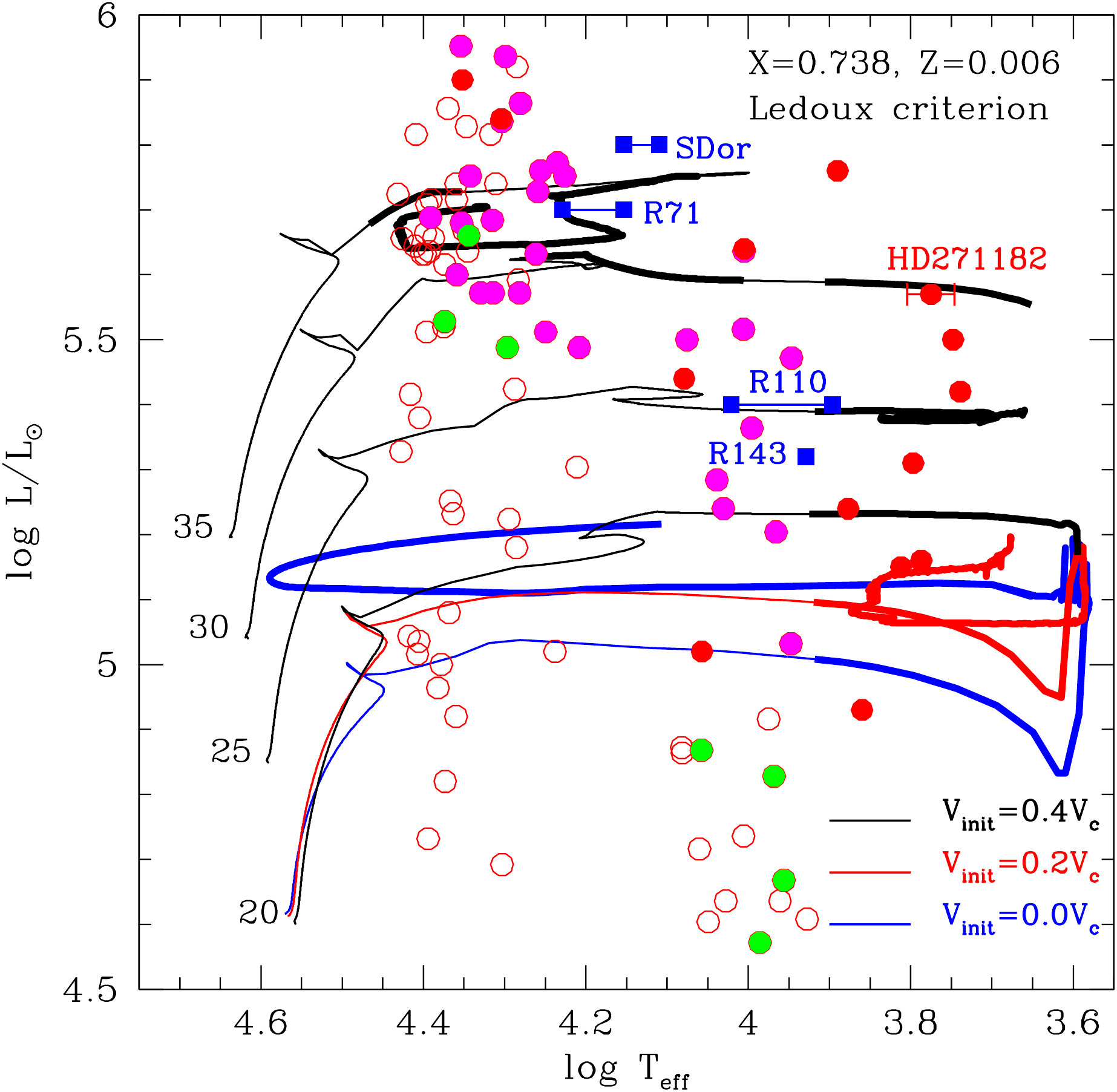}
\end{center}
\caption{Evolutionary tracks in the HRD for $Z=0.006$ with an initial rotation velocity of $0.4$ times critical value (black lines), of 0.2 times the critical value (red line), and without rotation (blue line). The left and right panels are for models with the Schwarzschild criterion and the Ledoux criterion, respectively. Thick lines indicate the parts of the tracks where radial pulsations are excited. Also shown are blue supergiants (luminosity classes of Ia, Iab) in the LMC: blue squares connected with horizontal lines are LBVs (S Dor, R 71, R 110, R 143; from brighter to fainter), for which parameters are obtained from \citet{sta90, lam98, vang01}. Red filled circles are known LMC $\alpha$ Cyg variables, whose parameters are given in Table~\ref{tab:var}. The other circles show the positions of BSGs of which parameters were obtained  by \citet{Urbaneja2017}. The photometric variability of each of these BSGs has been examined using the Fourier analysis software PERIOD04  \citep{period04} for the G-band lightcurve data from the ASAS-SN database \citep{Shappee2014a,Jayasinghe2019a}. Based on the analysis we show non-variables by open circles, probable $\alpha$ Cyg variables (with clear periodicities shorter than $\sim 100$ days) by filled magenta circles, and marginally variables by filled green circles.}
\label{fig:hrd_LMC}
\end{figure*}

Fig.~\ref{fig:hrd_LMC} compares the sections of the theoretical tracks where radial pulsations are excited (thick lines) with the observed positions of (semi-) periodic supergiant variables in the LMC (filled symbols). We make the comparison on the HR diagram (rather than the spectroscopic one) because the accurate knowledge of the distance to the LMC make the uncertainties in $\log L$ probably smaller than the uncertainties in $\log g$. Black tracks are for rotating models started with a rotation velocity of $0.4V_\text{c}$. For the case of  the Ledoux criterion (right panel), the non-rotating (blue line) $20\,M_\odot$ model and a rotating model with an initial velocity of $0.2 V_\text{c}$ (red line) are also shown for comparison. On the blue loop of the non-rotating model, radial pulsations are excited because it has lost a significant amount of mass in the previous RSG stage. The excitation of pulsations on the loop seems to contradict the absence of blue variables around the corresponding luminosity. However, due to the sensitivity of the loop to the rotation rate, an initial rotation rate of 20\% of the critical rate reduces significantly the loop, removing the discrepancy with the distribution of variable stars on the HR diagram. Since most of the massive stars rotate quite rapidly at this metallicity \citep[most of observed massive stars in the LMC have an equatorial velocity above $100\,\text{km}/\text{s}$,][]{RamirezAgudelo2013a}, we do not consider the long blue loop of the non-rotating model to be a shortcoming of the Ledoux criterion. We consider models with an initial rotation speed above about 20\% the critical one to be representative models to be compared with the properties of stars in the LMC: 20\% the critical velocity corresponds to an average equatorial velocity of about $125-150\,\text{km}/\text{s}$ during the MS.

\begin{figure*}
\includegraphics[width=0.49\textwidth]{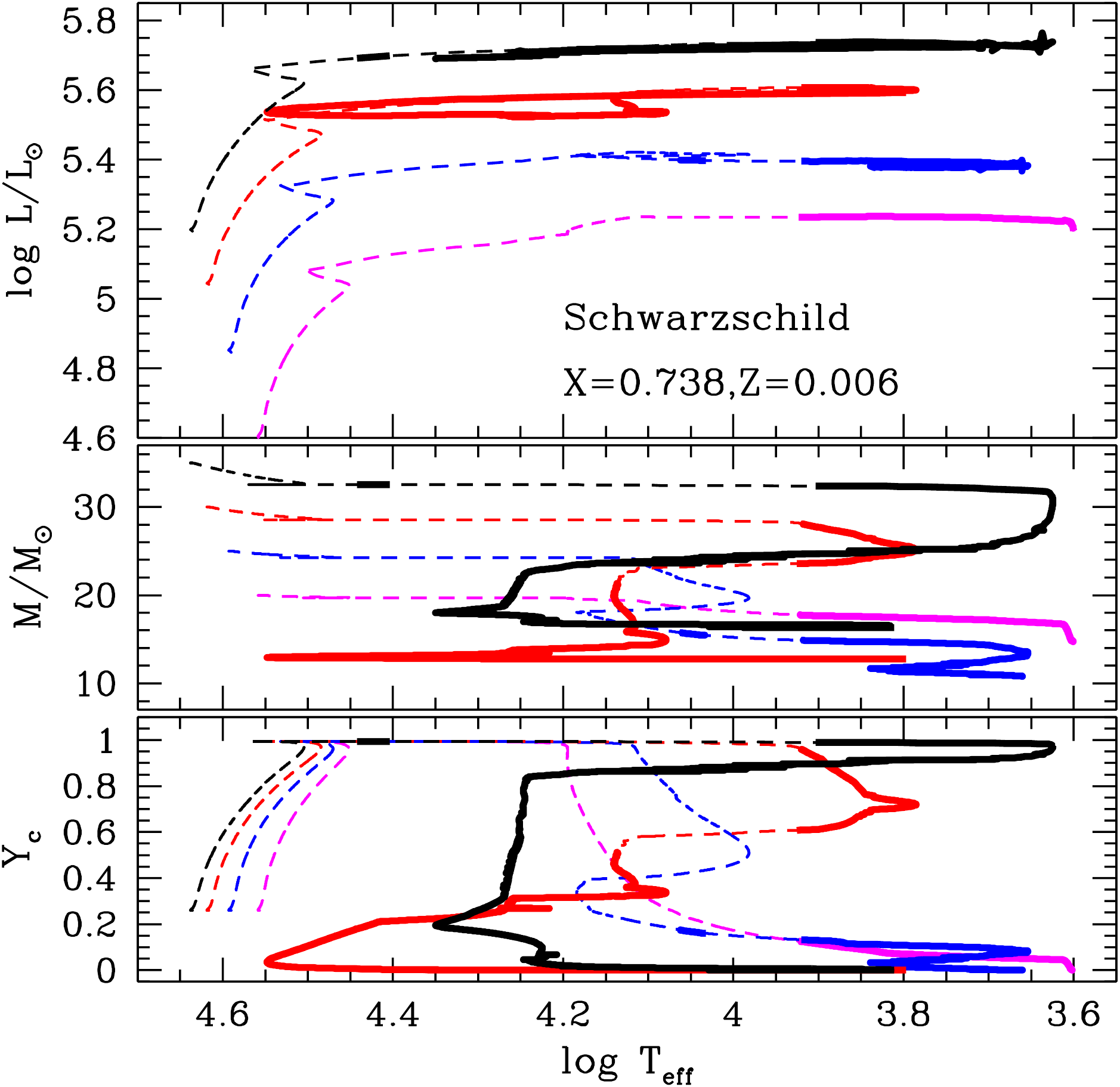}
\includegraphics[width=0.49\textwidth]{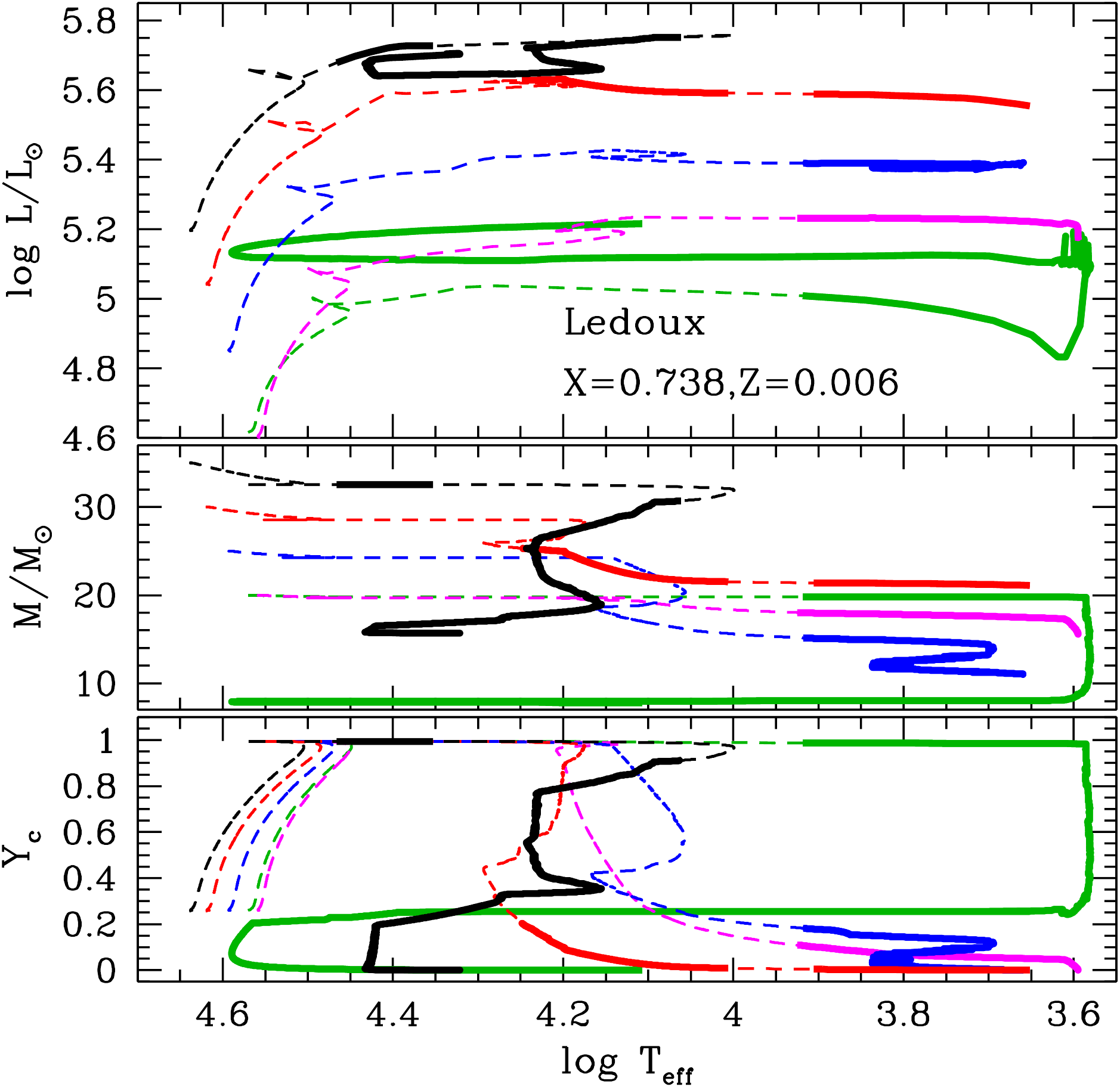}
\caption{Models with the LMC composition are shown in the same way as Fig.~\ref{fig:te_Lmyc}. The initial masses are: $20$ (magenta), $25$ (blue), $30$ (red) and $35\,M_\odot$ (black). Left panel shows models computed with the Schwarzschild criterion, and the right one models computed with the Ledoux criterion. The green line shows non-rotating model of $20\,M_\odot$ with the Ledoux criterion, while other cases are rotating models with an initial rotation of $0.4\,V_\text{c}$.
}
\label{fig:te_Lmyc_LMC}
\end{figure*}

\begin{table}
\caption{Selected non-LBV supergiant variables in the LMC}
\begin{center}
\begin{tabular}{llcll}
\hline
HD & $\log T_\text{eff}$ & $\log L/L_\sun$\tablefootmark{a} & Periods (d) & ref \\
\hline
 270920  &    3.748   &   5.50   &   250   & 1, 2  \\
 269018  &    4.057    &   5.02   & 14.6,  6.3, 32.5, 30.2  & 1, 2  \\ 
 33579    &    3.890   &   5.76   &  105,   81, 57, 70  & 1, 2 \\  
 271182  &    3.775    &   5.57   &  260   & 1, 2 \\ 
 269541  &    3.877    &   5.24   &  8.1, 24.6,12.0, 40.5  & 1, 2 \\
 269594  &    3.787    &   5.16   &  200  & 1, 2 \\
 269660  &    4.304    &   5.84   &  3.65, 10.8   & 1, 3 \\
 269697  &    3.797    &   5.31   &  48, 84  & 1, 2 \\
 269781  &    4.005    &  5.64   &  39.0    & 1, 3 \\
 268835  &  4.079       &   5.44   &   60,  380 & 4, 5 \\
 37974    &  4.352        &   5.90   &   24      & 4, 6 \\
 268757  &  3.739        &   5.42   &   540    & 7 \\
 268822  &  3.812        &   5.15   &   180    & 7 \\
 269612  &  3.86        &   4.93   &   $>100$  & 7 \\
\hline 
\end{tabular}
\end{center}
\tablefoot{\tablefoottext{a}{Luminosities were derived from $V$ magnitudes, assuming a mean extinction of $0.3$ mag and a distance modulus of $18.50$ mag, and adopting bolometric corrections from \citet{flo96}.}}
\tablebib{
(1) \citet{vanL98}; (2) \citet{mcd12}; (3) \citet{Urbaneja2017}; (4) \citet{vang02}; (5) \citet{sta83}; (6) \citet{zic85}; (7) \citet{vang04}
}
\label{tab:var}
\end{table}%

In Fig.~\ref{fig:hrd_LMC}, red filled circles are known LMC $\alpha$ Cyg variables whose parameters are given in Table~\ref{tab:var}, while the other circles are BSGs in the LMC whose  parameters were obtained  by \citet{Urbaneja2017}. We have examined the variabilities of these BSGs using the Fourier analysis software PERIOD04 \citep{period04} for the G-band lightcurve data from the ASAS-SN database \citep{Shappee2014a,Jayasinghe2019a}. Open circles are non-variables, filled magenta circles are probable $\alpha$ Cyg variables showing clear periodicities shorter than $\sim 100$ days, and filled green circles are stars which marginally show signs of variability. Squares connected with horizontal line are luminous blue variables (LBV), which are included as $\alpha$ Cyg variables, because it is known that the micro-variabilities of LBVs are $\alpha$ Cyg type variations caused by stellar pulsations \citep[e.g.][]{lam98}.

In this figure we see that  most of all variables are located either at high luminosity or in a cool enough location of the HRD ($\log T_\text{eff} < 3.9$), roughly agreeing with the theoretical prediction for the excitation of radial pulsations irrespective to the convective criteria employed. For more luminous and hotter stars which have lost considerable mass (see Fig.~\ref{fig:te_Lmyc_LMC}), the pulsations are excited by strange mode instability, while pulsations in  less luminous cool stars are excited mainly by the $\kappa$-mechanism as in the classical cepheids.

In contrast to our agreement of the LBV micro-variabilities with theoretical prediction of pulsational instability, \citet{lam98} claimed that the microvariations of the LBVs could not be explained by strange-mode instabilities, by comparing models obtained by \citet{kir93} who predicted instability in a region of $\log L/L_\sun > 6.0$ for $Z = 0.004$. The discrepancy can be attributed to the difference in the evolutionary models; our models include the effect of rotational mixing which increases the luminosity, and loses a considerable mass during the core-helium burning stage around $\log T_\text{eff} \sim 4.2$ (Fig.~\ref{fig:te_Lmyc_LMC}). Both effects increase the luminosity to mass ratio, which in turn enhances the effect of the strange mode instability. The explanation is consistent with the fact that \citet{lov14}'s result based on the old set of Geneva models \citep{mey94} without rotation indicates that the luminosity at the stability boundary is located between \citet{kir93}'s and ours.  

In summary, the distribution in the HRD of the LMC variable/non-variable supergiants does not indicate preference between the Ledoux and the Schwarzschild criteria.

\subsection{Surface compositions}

\begin{figure*}
\begin{center}
\includegraphics[width=0.49\textwidth]{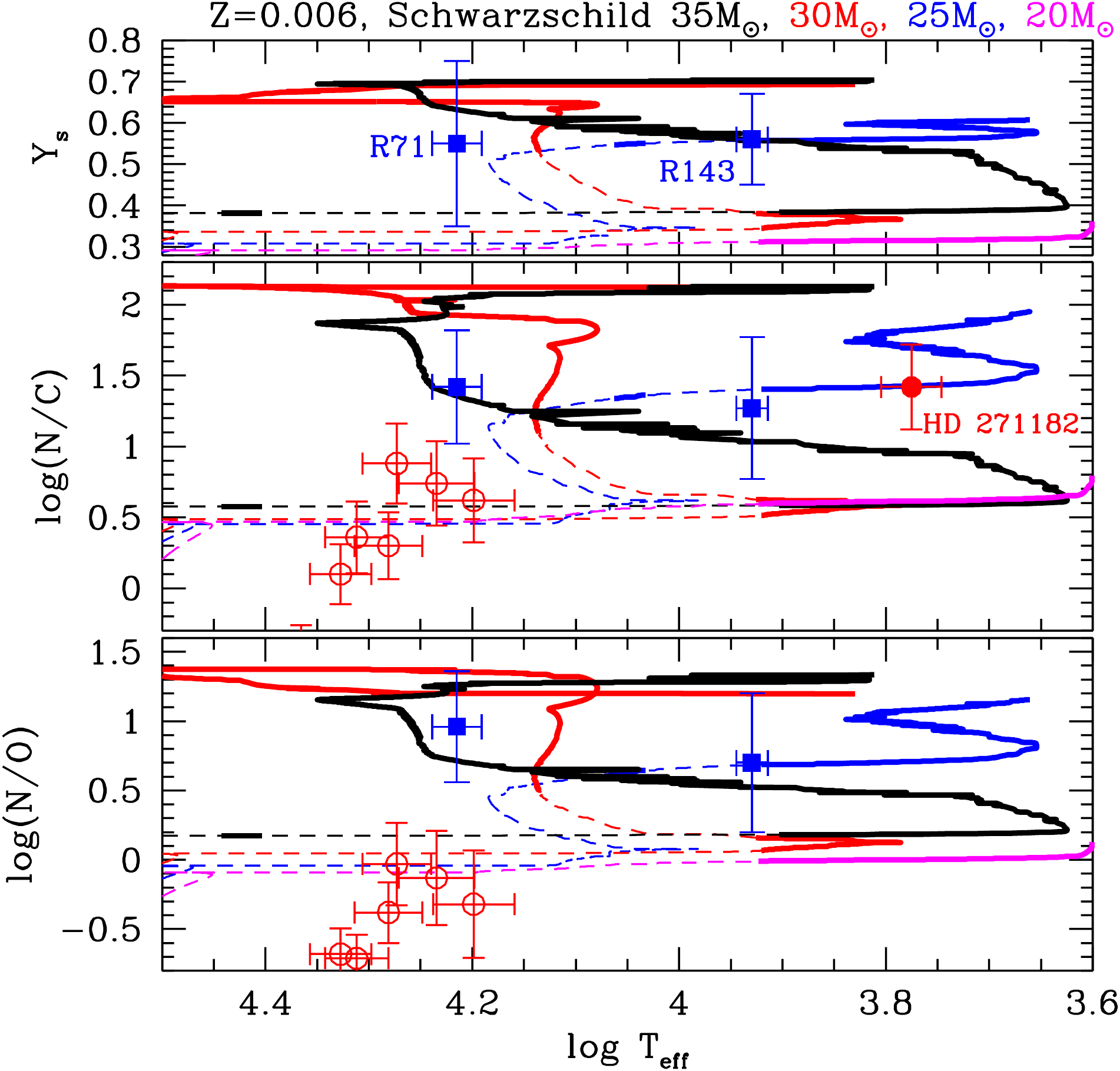}
\includegraphics[width=0.49\textwidth]{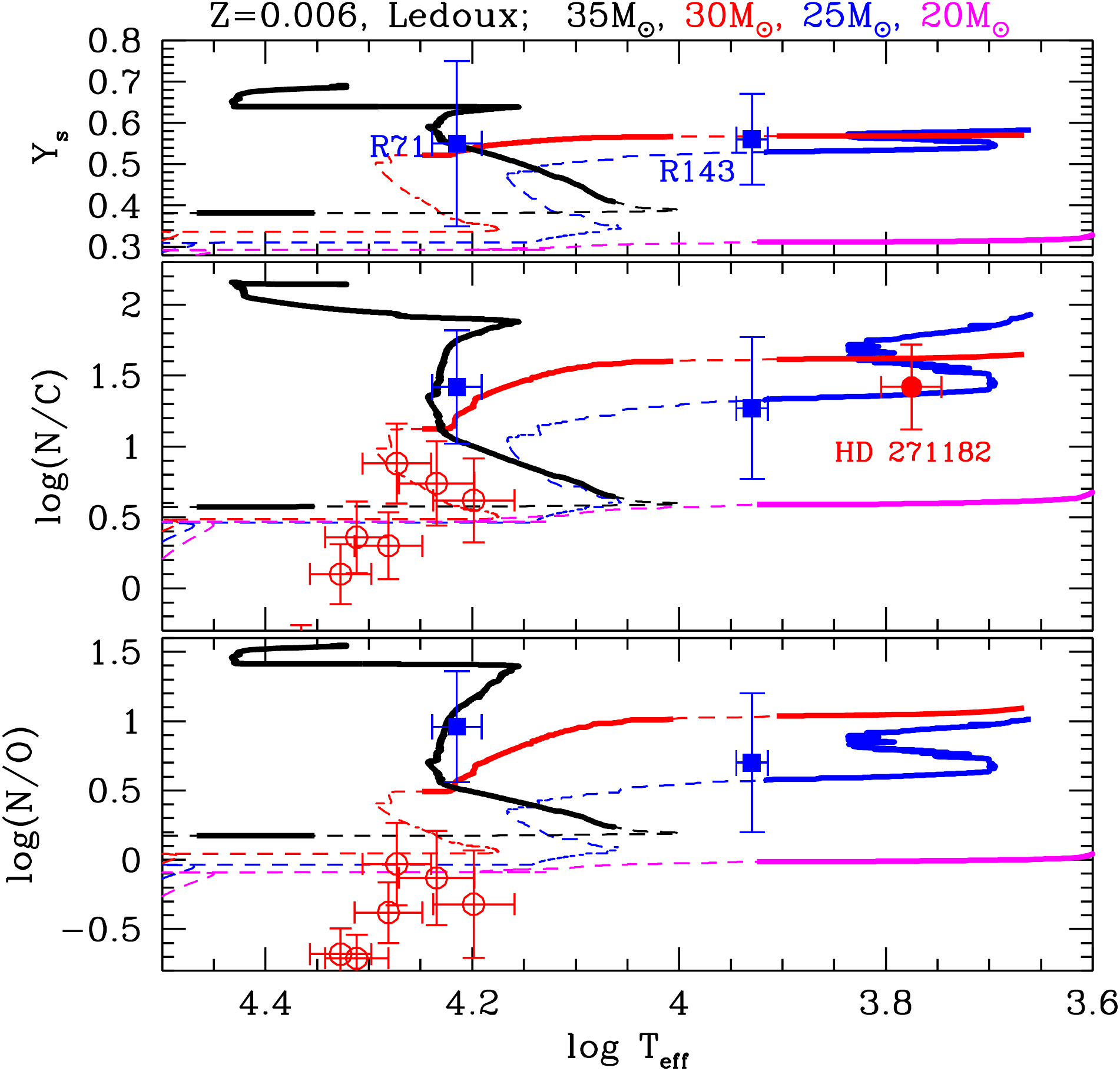}
\end{center}
\caption{Evolution of surface chemical compositions of $20$, $25$, $30$, and $35\,M_\sun$ models (magenta, blue, red, and black lines, respectively) with the Schwarzschild (left panel) and the Ledoux (right panel) criterion is presented as functions of $\log T_\text{eff}$. In those models the initial rotation speed is set to be 40\% of the critical value ($V_\text{c}$). Observational results available in the literature for massive LMC stars are also plotted; the meanings of the symbols are the same as in Fig.~\ref{fig:hrd_LMC}. The chemical compositions of the LBVs (filled squares) are adopted from \citet{len93} and \citet{Mehner2017a} for R~71, and \citet{Agliozzo2019} for R~143. The N/C ratio of the cool $\alpha$ Cyg variable HD~271182 is adopted from \citet{luc92}. Open circles are N/C and N/O ratios of blue supergiants (Ia, Iab) obtained by \citet{Hunter2009}; binaries are removed \citep{Maeder2014}.}
\label{fig:cno_LMC}
\end{figure*}

Fig.~\ref{fig:cno_LMC} shows the variations of surface helium abundance and CNO ratios during evolutions of $20$, $25$, $30$, and $35\,M_\sun$ (magenta, blue, red, and black lines, respectively) with the Schwarzschild (left panel) and the Ledoux (right panel) criteria. The initial rotation speed is assumed to be 40\% of the critical speed ($V_\text{c}$). Thick solid lines represent evolutionary stages when radial pulsations are excited, while on the thin dashed line parts no radial pulsations are excited. After the main sequence evolution, in which rotational mixing modifies surface composition, the surface composition start changing at the beginning of  the core-helium burning for models massive enough so that the mass loss is significant (the models of $20\,M_\sun$ hardly change the surface composition after the end of main sequence, because they lose only a little amount of mass; see Fig.~\ref{fig:te_Lmyc_LMC}). The increase in the surface N/C and N/O ratios and helium abundances are comparable in both cases of the Ledoux and Schwarzschild  criteria, in contrast to the case of solar metallicity shown in Fig.~\ref{fig:CNOys}.

Kippennhahn diagrams are shown in Fig.~\ref{fig:KippenAll} for the $25\,M_\sun$ model with Schwarzschild criterion (bottom left panel) and the one with the Ledoux criterion (bottom right panel). There is no drastic difference in the extension of the convective shell between the two cases, in contrast to the solar metallicity models (top panels of Fig.~\ref{fig:KippenAll}). In both cases, the outer boundary of the shell convection zone extends to about $M_r \approx 18\,M_\odot$ at the contraction stage after the main sequence evolution. Due to wind mass loss during the core-helium burning stage, the stellar surface reaches the layers where material was previously processed by H-burning. Then the surface N/C and N/O ratios increase steeply. Since the maximum extent of the convective shell is comparable between the models with the Schwarzschild and the Ledoux criterion, the amounts of changes in the surface compositions are comparable in the two cases for models of the LMC composition. 

It is not easy to explain the changes in the behaviours of the intermediate H-burning shell at various metallicities and for different criterion for convection. An important point is whether the model starts core-helium burning in the blue part of the HRD or crosses the HRD and starts core-helium burning as a RSG. This depends on a variety of parameters, which can be interdependent, such as : rotation (and its implementation), mass-loss rates, chemical gradients in the radiative zones, activity of the intermediate H-burning shell, etc. \citep{Langer1995a,Maeder2001a}. Models at low metallicity tend to remain longer in the blue region of the HRD \citep[e.g.][]{Georgy2013b}, and this is also what we observe in the computations used in this work. A similarity between our models at both metallicity is that the luminosity of the hydrogen burning-shell at the onset of shell hydrogen-burning is systematically higher for models computed with the Schwarzschild criterion compared to models computed with the Ledoux one (see Fig.~\ref{fig:LumShell}). This is due to chemical gradients in the region which was previously occupied by the hydrogen burning-core during the MS (the region below the dot-dashed line in Fig.~\ref{fig:KippenAll}), which prevent a convective zone to appear at the same location where hydrogen is burnt. Hydrogen-shell burning starting in radiative condition with the Ledoux criterion, there is no refueling in fresh hydrogen by convection and then the energy production remains lower. In the Ledoux models, the convective zone appears higher inside the star, where no burning occurs. It then growth at deeper level by eroding the chemical gradient below the convective zone.

The difference between solar and LMC metallicity is that at solar metallicity, the first crossing of the HRD can be fast: all the Ledoux models and the Schwarzschild $25\,M_\sun$ models ignite central helium burning on the RSG branch after a quick crossing of the HRD. Only the Schwarzschild $20\,M_\sun$ model starts burning its helium as a BSG before slowly crossing the HRD for the first time: the luminosity of the intermediate hydrogen-burning shell is sufficient to maintain the model on the blue side of the HRD, and remains about constant, while in other cases, it decreases over time. Figure~\ref{fig:LumShell} shows the time-evolution of the shell luminosity for a couple of $10^5$ years after the end of the MS. Three different kinds of behaviour can be seen: 1) models with an abrupt decrease in the shell luminosity short after the shell ignites (Ledoux models at solar metallicity). These are the models which are crossing the Hertzsprung gap very quickly after the end of the MS. 2) models with a slow decrease of the shell luminosity over time (models at $Z=0.006$ and Schwarzschild solar metallicity $20\,M_\sun$ model). These models ignite core-He burning in the blue part of the HRD and cross the gap slowly. 3) model with an intermediate behaviour (Schwarzschild solar metallicity $25\,M_\sun$ model). This model is crossing the HRD quite quickly, but not as fast as the models listed in 1). For the models that are quickly crossing the HRD, the intermediate convective zone of the Ledoux models have no time to erode the chemical gradients and to reach the same position as in the Schwarzschild case before the expansion of the external layers switches the intermediate convective zone off. The regions mixed up by convection are thus located at different depth inside the star at this metallicity, changing the surface chemical composition later on, when mass loss uncovers these regions.

At lower metallicity, the luminosity of the hydrogen-burning shell is still higher in the Schwarzschild than in the Ledoux models (see Fig.~\ref{fig:LumShell}). However this difference is no longer enough for making the Ledoux model crossing quickly the HRD. At this metallicity, core helium-burning starts in the blue part of the HRD in both cases. The convective shell of the Ledoux model appears at a higher level inside the star as shown in Fig.~\ref{fig:KippenAll} (bottom right panel). Then convection erodes the chemical gradient below the convective zone, which progressively reaches the same position as in the Schwarzschild model, preventing a drop in the luminosity of the intermediate burning shell of the Ledoux models, as can be seen in Fig.~\ref{fig:LumShell}. The external layers of the star does not expand, keeping the star at quite high effective temperature, and the conditions for keeping an active intermediate convective shell are preserved, contrarily to the solar metallicity case. It leads to a very similar evolution of the intermediate convective shell, independent of the criterion used for convection.

Spectroscopic surface helium abundance, $Y_\text{s}$ and CNO ratios of LMC supergiants available in the literature are plotted in Fig.~\ref{fig:cno_LMC}. For non-variable supergiants (Ia, Iab), results from the VLT-FLAMES survey \citep{Hunter2009} are shown; in their analysis the He/H number ratio assumes to be $0.1$ (corresponding to $Y_\text{s} = 0.285$). These blue supergiants seem to be on the evolution stage just after the main sequence before helium ignition so that the surface N/C and N/O ratios reflect the rotational mixing during the main-sequence stage (we note that if we adopt the Ledoux criterion two stars with highest N/C ratios could be in the core-helium burning stage according to the $30\,M_\sun$ model). The N/C ratios of these blue supergiants look comparable (except for one extremely deficient one) with model predictions, while the theoretical predictions of N/O ratios tend to be larger than the observed ratios \citep[see discussion in][]{Maeder2014}.

Very limited spectroscopic results for variable supergiants in the LMC are available; plotted in Fig.~\ref{fig:cno_LMC} are $Y_\text{s}$ and N/C, N/O ratios for the LBVs R~71 and R~143 (filled squares) obtained by \citet{len93} and \citet{Mehner2017a,Agliozzo2019}, respectively, and the N/C ratio of the cool $\alpha$-Cyg variable HD~271182 (filled circle) obtained by \citet{luc92}. 

On the HR diagram R~71 and R~143 are located close to the $35\,M_\sun$ and $25\,M_\sun$ tracks (Fig.~\ref{fig:hrd_LMC}), corresponding to black and blue lines, in Fig.~\ref{fig:cno_LMC} respectively. This figure indicates that the helium abundances and CNO abundance ratios of R~71 and R~143 are roughly consistent with models either with the Ledoux or Schwarzschild criterion.

For the cool $\alpha$-Cyg variable HD~271182, only the N/C ratio obtained by \citet{luc92} is available. On the HR diagram (Fig.~\ref{fig:hrd_LMC}), HD~271182, which is shown by a filled circle with horizontal error bars, is located along the $30\,M_\sun$ track in either case of Ledoux or Schwarzschild criterion. The surface N/C and the position on the HRD of HD~271182 are consistent only with the model with the Ledoux criterion, but inconsistent with the Schwarzschild criterion. This supports, although only weakly, the Ledoux criterion for the LMC composition. Recently, \citet{Neugent2012} identified many ($\sim300$) yellow supergiants separating from dominant foreground Galactic dwarfs. We hope that spectroscopic abundance analyses as well as time resolved photometries will be done for these yellow supergiants (luminous ones in particular) in a near future.

To summarize, the properties of BSGs in the LMC do not indicate a clear preference between the Schwarzschild and the Ledoux criterion. In case the evolutionary mass of HD~271182 is correct (about $\sim30\,M_\sun$), then the Ledoux criterion is however better in reproducing the observed characteristics of this star in particular.

\section{Flux-weighted gravity--Luminosity Relation}

\begin{figure}
\includegraphics[width=0.49\textwidth]{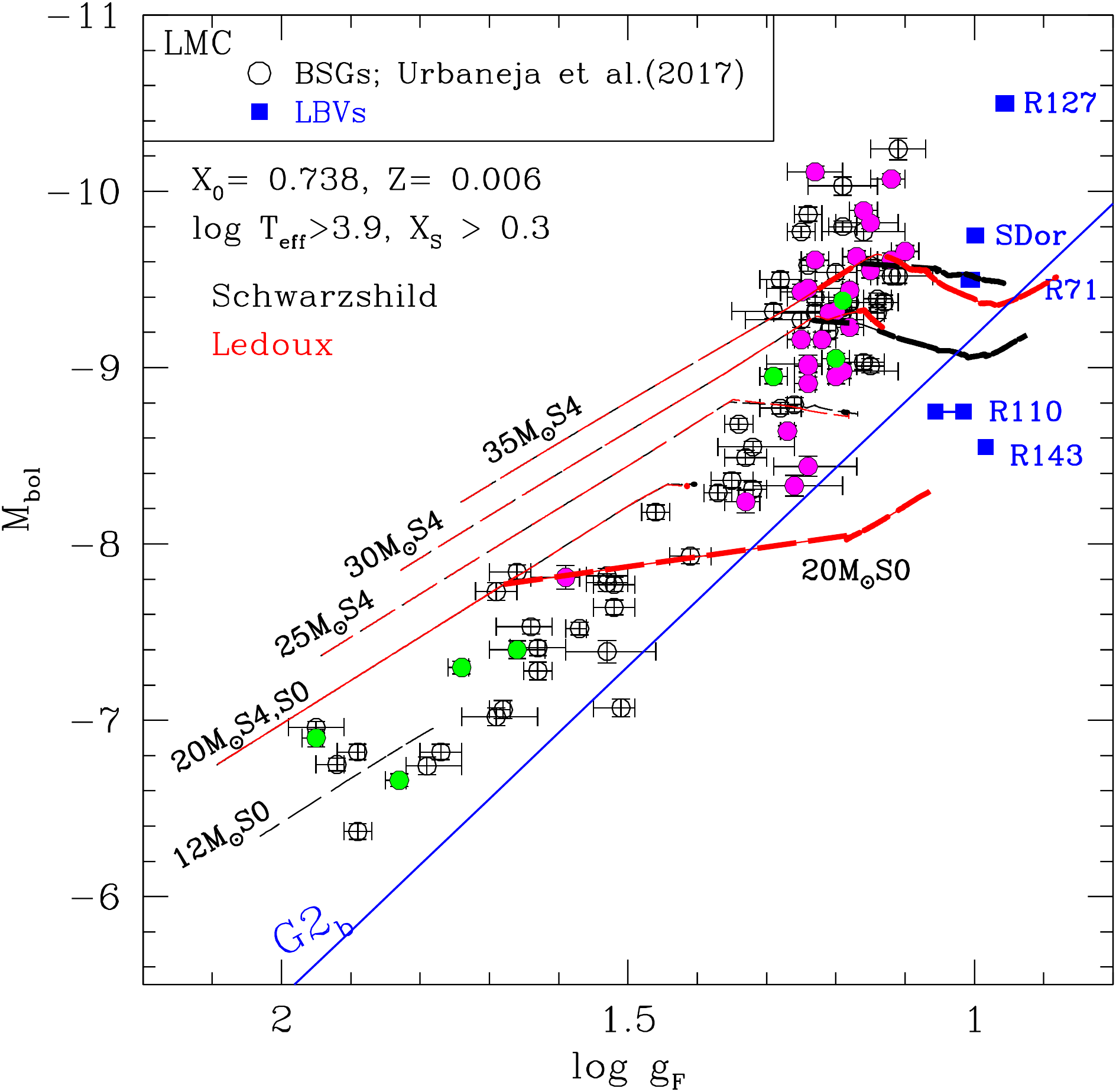}
\caption{The LMC blue supergiants analysed by \citet{Urbaneja2017} are plotted by circles in the $\log g_\text{F} - M_\text{bol}$ plane. The colour of the circles have the same meaning as in Fig.~\ref{fig:hrd_LMC}. The blue straight line labeled as G2$_b$ is the lower boundary of the FGLR of BSGs in Local group galaxies \citep{Meynet2015}. Blue parts ($\log T_\text{eff} > 3.9$) of some evolutionary tracks with $Z=0.006$ are also plotted, where thick line parts indicate where radial pulsations are excited. The labels ``S0'' and ``S4'' refer to models with an initial rotation rate of $0$ and $0.4$ of the critical one respectively. Some LBVs are plotted by filled squares for comparison, which clearly indicate that LBVs have experienced significant mass losses.
}
\label{fig:FGLR_LMC}
\end{figure}

The Flux-weighted gravity--Luminosity Relation was introduced by \citet{Kudri2003,Kudri2012} to spectroscopically measure the distances to galaxies. Later, \citet{Meynet2015} has shown that the relation should be a powerful tool to discriminate between BSG1 and BSG2 because the relation depends on the mass-loss occurred during the stellar evolution. It is thus an important check for our models to compare the predictions of our computations with the observed position of variable BSG in this diagram. The flux-weighted gravities $g_\text{F}$, defined as
\begin{equation}
g_\text{F} = \frac{g}{(T_\text{eff}/10^4\,\text{K})^4},
\end{equation}
of BSGs were found by \citet{Kudri2003} to form a tight relation with luminosity (or absolute bolometric magnitude, $M_\text{bol}$), where $g$ is the surface gravity and $T_\text{eff}$ the effective temperature. A most recent calibration is obtained from detailed spectroscopic analysis of blue supergiants in the LMC \citep{Urbaneja2017}. Figure~\ref{fig:FGLR_LMC} shows the relation with some evolutionary tracks at $Z=0.006$. The blue line labeled G2$_b$ is the relation corresponding to $\log L/L_\sun = 3\log M/M_\sun + 2.03$ obtained by \citet{Meynet2015} as the lower boundary of the FGLR for BSGs of Local Group galaxies. This figure indicates that the FGLR of LMC is also bounded by the same relation as that of Local Group galaxies.

The positions of the LBVs (filled blue squares) in the $g_\text{F}-M_\text{bol}$ plane are separated from the mean FGLR towards  the lower $g_\text{F}$ side. Since $g_\text{F}\propto M/L$, the LBV positions indicate that they have lost significantly more mass than the ordinary BSGs, which is consistent with our common understanding of the LBVs. In contrast to the fact that the LBV positions in the HRD are intermingled with the ordinary BSGs (Fig.\,\ref{fig:hrd_LMC}), the segregation of LBVs in the $g_\text{F}-M_\text{bol}$ plane is remarkable, and could be useful for finding LBV candidates.

\citet{Meynet2015} discussed on the consistency of the FGLR of Local Group low metallicity galaxies with the theoretical evolution models of \citet{Ekstrom2012}. According to the evolution models, stars whose initial masses are larger than $\sim 20\,M_\sun$ become BSG for a second time after considerable mass is lost in the red supergiant stage. The BSG2 would be located right side of the G2$_b$ line, and \citet{Meynet2015} concluded that the tightness of the FGLR indicates that the evolution towards group 2 BSGs should be rare in local group galaxies. This conclusion is somewhat inconsistent with our identification of $\alpha$ Cyg variables as BSG2s \citep{Saio2013}.

\begin{figure*}
\includegraphics[width=0.49\textwidth]{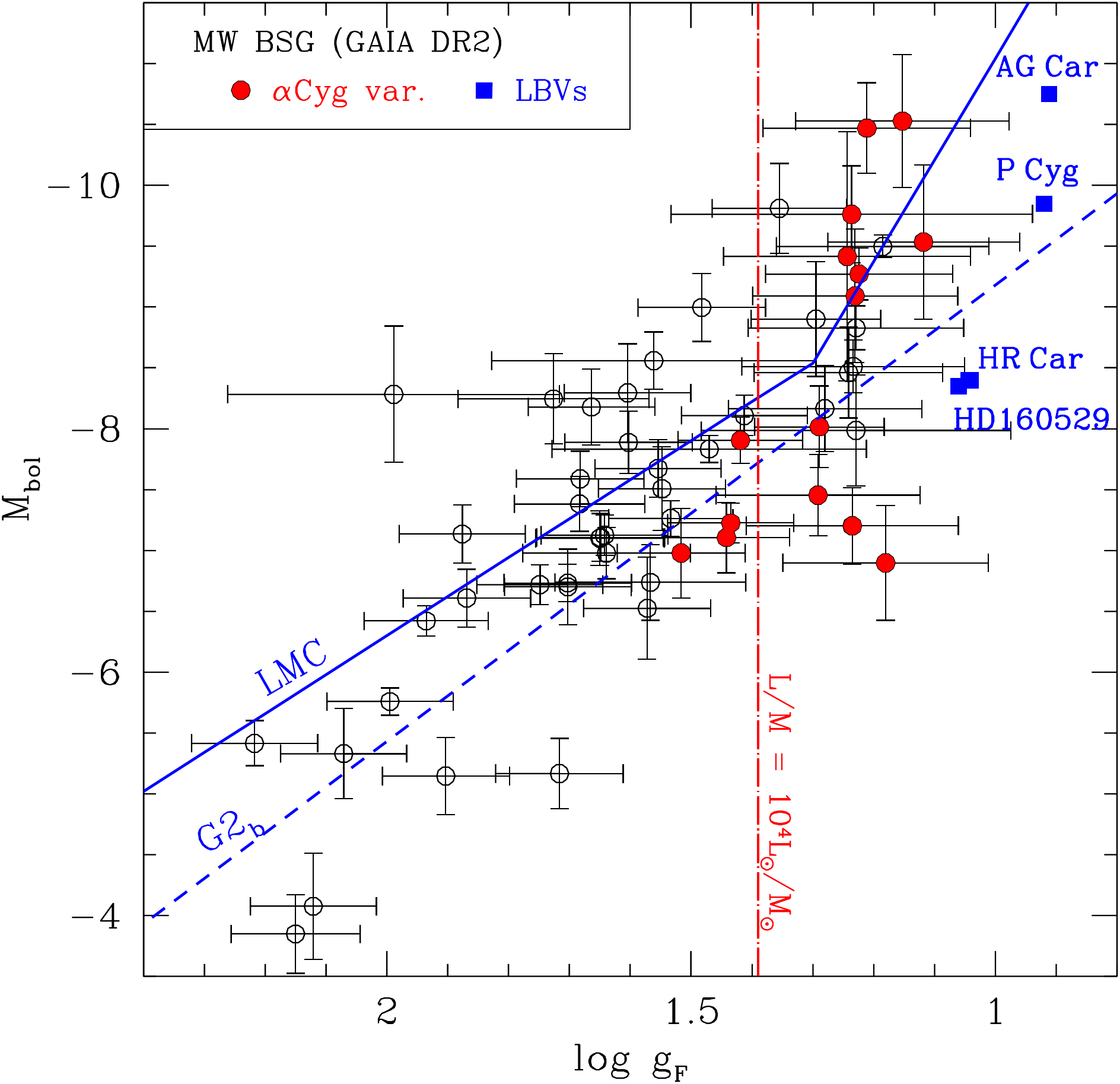} 
\includegraphics[width=0.49\textwidth]{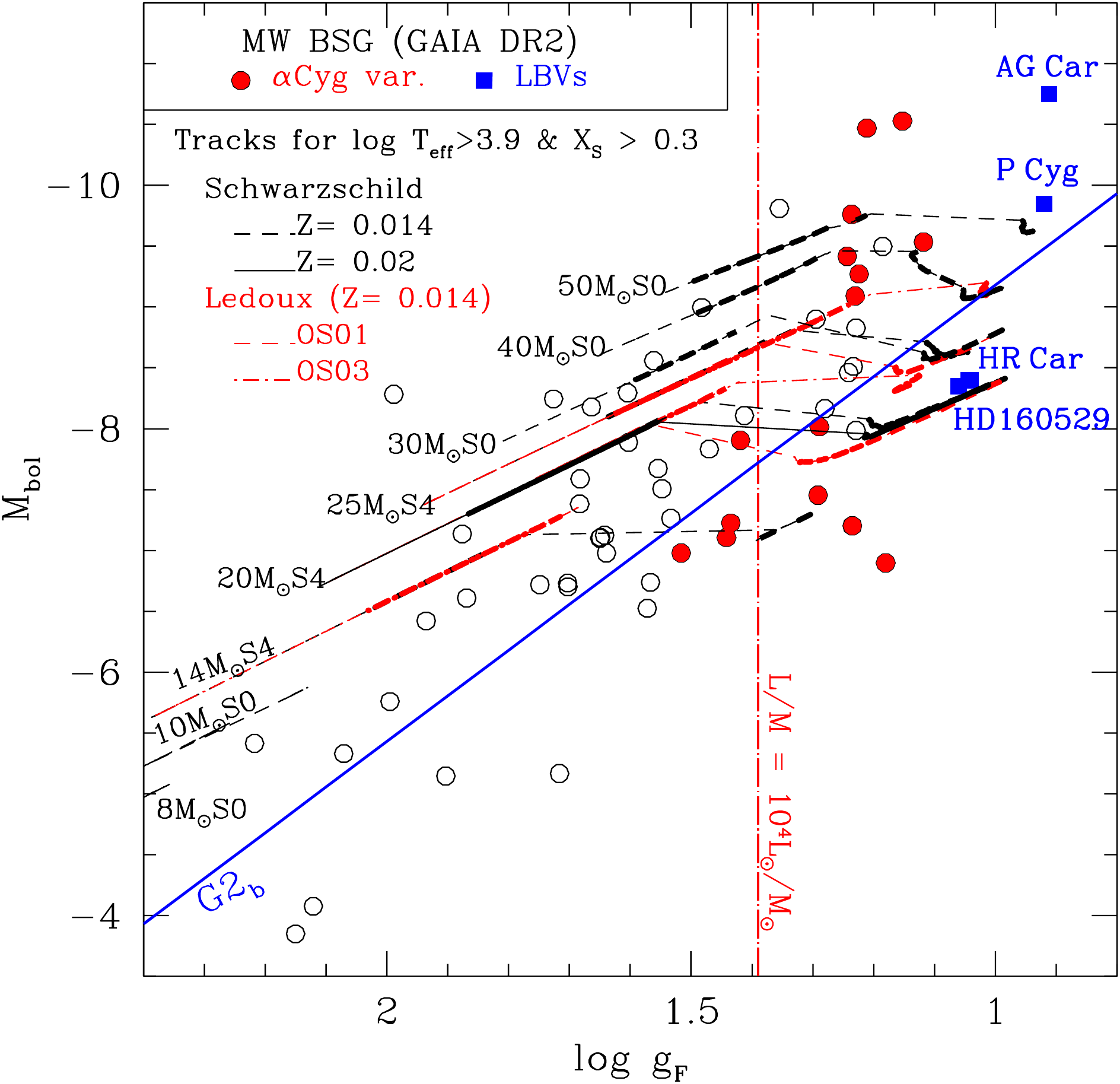} 
\caption{\textit{Left:} Absolute bolometric magnitudes $M_\text{bol}$ versus flux-weighted gravity $g_\text{F}$ of Galactic blue supergiants are plotted with error bars. For most of the stars GAIA DR2 parallaxes \citep{Gaia2018b} are used. A blue line indicates the mean relation of LMC blue supergiants obtained by \citet{Urbaneja2017}. Blue dashed line labeled G2$_b$ corresponds to the relation $\log L/L_\sun = 3\log M/M_\sun +2.03$ \citep{Meynet2015} which gives the lower bound of the FGLR for the BSGs of Local Group galaxies. \textit{Right:} Blue parts ($\log T_\text{eff} > 3.9$) of evolutionary tracks with various parameters are compared. The labels ``S0'' and ``S4'' refer to models with an initial rotation rate of $0$ and $0.4$ of the critical one respectively. Some Galactic LBVs are shown by filled blue squares. The loci on the $g_\text{F}-M_\text{bol}$ diagram clearly indicate significant mass losses have occurred in LBVs.
}
\label{fig:FGLR}
\end{figure*}

The conclusion of \citet{Meynet2015} is based on the FGLR of the BSGs in Local Group galaxies. It is more desirable to examine the theoretical and observational consistency using BSGs in our Galaxy. Thanks to the recent second release of the GAIA parallax data, DR2 \citep{Gaia2016,Gaia2018b}, it is now possible to accurately plot Galactic blue supergiants on the $g_\text{F}-M_\text{bol}$ plane (Fig.~\ref{fig:FGLR}). The spectroscopic data of blue supergiants of the Galaxy (see Table~\ref{tab:sum}) needed to obtain $g_\text{F}$ are adopted from the literature \citep{Crowther2006,Searle2008,Przy2010,Firnstein2012,Clark2012}. Among the plotted BSG stars, known $\alpha$ Cyg variables are shown by filled circles. For the identifications of the $\alpha$ Cyg type variability, we consulted the literature based on the Hipparcos photometry  \citep{Koen2002,Lefevre2009,Dubath2011,Rimoldini2012}.

As seen in Figure~\ref{fig:FGLR}, the FGLR of the Galactic blue supergiants are similar to that of the LMC, but possibly more extended to the lower $g_\text{F}$ side, which may indicate that the blue supergiants in the Galaxy experienced more mass loss than the blue supergiants in the LMC having similar luminosity. This is consistent with the evolution tracks of rotating stars  presented by \citet{Ekstrom2012}. Also, it is interesting to note that the loci of the Galactic LBVs in the $g_\text{F}-M_\text{bol}$ plane (filled blue squares in Fig.~\ref{fig:FGLR}) are similar ($g_\text{F}\sim10\,\text{cm}\,\text{s}^{-2}\,(\text{K}/10^4)^4$) to those of the LMC LBVs (Fig.~\ref{fig:FGLR_LMC}) despite the large difference in the metallicity.

In the $g_\text{F}-M_\text{bol}$ plane, the BSG2s should be located systematically on the lower side $g_\text{F}$ side for a given $M_\text{bol}$ compared with the location of BSG1s. The parts of evolution tracks for the BSG2s can be recognized in the right panel of Figure~\ref{fig:FGLR} as the (gently ascending) thick-line parts below the G2b line where pulsations are excited. In addition, this figure indicates that radial pulsations are also excited well above the G2b line if $M \ge14 M_\sun$. This is caused by the $\kappa$-mechanism in the $\beta$-Cep instability strip. As the luminosity increases (i.e., $L/M$ increases), the effect of strange-mode instability gets stronger and the instability range widens to include very luminous $\alpha$ Cyg variables.

In fact, the Galactic $\alpha$-Cyg variables (red filled circles) are separated into relatively low and very high luminosity groups at $M_\text{bol}\sim -9$ (or $\log L/L_\sun \sim 5.5$) in the $g_\text{F}-M_\text{bol}$ plane. The majority of $\alpha$ Cyg variables with $M_\text{bol} > -9$ are below the G2b line in Figure~\ref{fig:FGLR}, indicating the relatively less luminous $\alpha$ Cyg variables are BSG2s, while there are some very luminous $\alpha$ Cyg variables, which belong to BSG1s.

Since $g_\text{F}\propto M/L$, a constant $L/M$ corresponds to a vertical line in the $g_\text{F}-M_\text{bol}$ diagram. The location of $L/M=10^4L_\sun/M_\sun$ is shown as a red line in Figure~\ref{fig:FGLR}. A star with $L/M > 10^4 L_\sun/M_\sun$ should fall in the right side of the vertical line. This figure shows that for the majority of $\alpha$ Cyg variables  have $L/M \gtrsim 10^4 L_\sun/M_\sun$, which indicates that radial pulsations of $\alpha$ Cyg variables are excited by the strange mode instability \citep[e.g.][]{Glatzel1994,Saio1998,Saio2009}\footnote{This also explains the fact that most of the variable LMC BSGs in Fig.\,\ref{fig:FGLR_LMC}  are located in the range $\log g_\text{F} < 1.4$.}.

There are some less luminous ($M_\text{bol} \gtrsim -6$) stars located below the G2$_b$ line. They cannot be explained by single star evolutions, indicating that transferring a significant envelope mass to a companion in a close binary system would be needed.

\section{Conclusion}

In this work, we compare models of massive stars computed with both the Schwarzschild and Ledoux criterion with observed pulsating BSGs ($\alpha$ Cyg variables). In particular, we used the position in the HRD, surface chemical composition, excitation of radial modes, and position in the flux-weighted gravity-luminosity diagram to test our models. Confirming our preliminary results \citep{Georgy2014}, our comparisons with observations show the Ledoux criterion is better than the Schwarzschild one at solar metallicity. It improves particularly the fit with the observed surface chemical abundances while keeping a good agreement with the location where radial pulsations are observed in the HRD. This also supports the idea that relatively less luminous ($\log L/L_\sun \lesssim 5.5$) Galactic $\alpha$ Cyg variables are group 2 BSGs, i.e. stars that have had a previous RSG stage before crossing the HRD for a second time towards higher effective temperatures. In that case, quite high mass-loss rates during the RSG phase is needed to favour bluewards evolution after the RSG phase.

However, our models still have difficulties in reproducing the surface helium abundance of observed stars, although models computed with the Ledoux criterion are closer to the observations. We also tried to change input physics for convection, by changing the efficiency of the overshoot at solar metallicity. We find that this does not help to improve the agreement between the models and the surface chemical abundances observed in $\alpha$ Cyg variables.

At the LMC metallicity, we show that models produce rather similar results independent of the chosen criterion for convection. This is due to the fact that at this metallicity, changing the criterion does not impact the location where helium-core burning starts in the HRD: it begins in the blue side of the HRD in both cases. Comparison with observations does not strongly favour any of the criterion at this metallicity.

We have compared the FGLR of LMC BSGs \citep{Urbaneja2017} with our evolutionary tracks. We find that variable BSGs are located in the range of $\log g_{\text F} \lesssim 1.4$ (i.e., $L/M \gtrsim 10^4 L_\sun/M_\sun$). The majority of the LMC BSGs form a relatively tight sequence, which indicates mass losses from them have not been very significant. In contrast, LBVs are located significantly lower $g_\text{F}$ side deviating from the tight relation of the other BSGs, which indicates that they have lost significant mass, in agreement with our common understanding of the LBVs.  

We also compare the FGLR of Galactic $\alpha$ Cyg variables with the results of our modeling. We find that the FGLR of Galactic BSGs is broader than that of LMC, indicating that wind mass-loss is more active. The FGLR clearly shows that relatively less luminous $\alpha$ Cyg variables are members of group 2 BSGs, while some very luminous $\alpha$ Cyg variables belong to group 1 BSGs.

Our findings suggest that the use of the Ledoux criterion for convection produce slightly better agreements with the observations. More work is needed to confirm this result, since it is probably not independent on other choices made in our models (mass-loss rates, implementation of rotation, implementation of convective boundary mixing, ...). We will continue our efforts to improve the constraints on stellar models that can be deduced from the comparison with observations of BSGs.

\begin{acknowledgements}
The authors thank the anonymous referee for her/his valuable comments that contributed to improve this work. This work has made use of data from the European Space Agency (ESA) mission \textit{Gaia} (\url{https://www.cosmos.esa.int/gaia}), processed by the \textit{Gaia} Data Processing and Analysis Consortium (DPAC, \url{https://www.cosmos.esa.int/web/gaia/dpac/consortium}). Funding for the DPAC
has been provided by national institutions, in particular the institutions participating in the \textit{Gaia} Multilateral Agreement. CG and GM have received funding from the European Research Council (ERC) under the European Union’s Horizon 2020 research and innovation  program  (Grant  Agreement  No.  833925).
\end{acknowledgements}

\bibliographystyle{aa}
\bibliography{Biblio}

\end{document}